# Bayesian sequential decision-making for adaptive early warning of bridge failure risk

Teng Li[a,b], James L. Beck[c], Yong Huang[a,b,*], Hui Li[a,b], Hongju Han[d]

[a] Key Lab of Smart Prevention and Mitigation of Civil Engineering Disasters of the Ministry of Industry and Information Technology, Harbin Institute of Technology, Harbin 150090, China

[b] Key Lab of Structures Dynamic Behavior and Control of the Ministry of Education, School of Civil Engineering, Harbin Institute of Technology, Harbin 150090, China

[c] Division of Engineering and Applied Science, California Institute of Technology, CA, USA

[d] Guizhou Key laboratory of Intelligent Construction and Maintenance of Mountain Bridge and Tunnel, Guiyang 550005, China

## Abstract

As a critical functionality of structural health monitoring (SHM) systems, early warning of increased bridge failure risk is essential for structural safety. However, traditional methods treat early warning merely as passive detection against a fixed safety limit and neglect evolving uncertainties and the economic consequences of decisions, leading to delayed alarms that increase the failure risk and hinder reliable and rational decision-making. This study formulates the early warning of increased structural failure risk as a sequential decision-making problem and develops a Bayesian framework for its solution. The presented framework unifies a probabilistic state transition model, an observation prediction model, cost models for early warning decisions, and a threshold-based decision strategy into a closed loop encompassing update, prediction, optimization, and decision execution. The structural state is updated iteratively via Bayes' theorem upon the arrival of new monitoring data, ensuring each decision is conditioned on the latest information. Through pre-posterior analysis over predicted future observations, the warning threshold is optimized by minimizing the expected total cost, which explicitly balances structural failure risks against warning response costs. In this way, the warning paradigm is shifted from passive detection to proactive decision-making, with the warning threshold adaptively updated as uncertainties and costs evolve. The framework was verified via two bridge case studies covering typical scenarios that lead to increased structural failure risk, including steel truss member fracture and stay cable rupture. Results demonstrate that warning decisions can be triggered in a timely and effective manner before severe consequences occur.

*Corresponding authors.

E-mails: huangyong@hit.edu.cn (Yong Huang)

## 1. Introduction

As a critical component of modern transportation networks, bridges play a vital role in economic development and social connectivity. However, bridge structures inevitably deteriorate throughout their service life. The failure of these structures typically leads to catastrophic consequences, resulting in significant economic losses and human casualties. To ensure structural safety, structural health monitoring (SHM) has been widely implemented on bridges [1,2], providing continuous observations of structural response through instrumented sensing systems. As a critical functionality of SHM, early warning systems can detect potential structural abnormalities from monitoring data at an early stage, enable timely interventions, prevent structural failures, and optimize the allocation of limited operation and maintenance resources. Consequently, developing intelligent early warning methods capable of assessing structural failure risk level and triggering timely warning decisions has become an urgent research priority in civil engineering [3-5].

The purpose of early warning for bridge failure risk is not merely to identify potential anomalies, but to support timely intervention. Deciding whether and when to take action, however, is far from straightforward. Bridge managers must choose among alternative actions, such as keeping the bridge in normal service or issuing an early warning and closing it to traffic (or other actions). The consequences of an inappropriate choice can be severe: a delayed warning exposes the public to the risk of catastrophic collapse, whereas an unnecessary warning disrupts traffic and wastes limited operation and maintenance resources. Benefiting from the development of SHM techniques [6-8], sequential monitoring methods have been introduced into civil engineering for early warning applications. These methods assess the structural condition or safety indicators based on data samples that arrive continuously over the service life, thereby deciding whether an alarm should be triggered. Early efforts borrowed control charts from statistical process control, including the Shewhart, exponentially weighted moving average (EWMA), and cumulative sum (CUSUM) schemes, in which an alarm is raised once the charting statistic of a damage-sensitive feature exceeds its control limit [9,10].

Multivariate control charts were later developed to fuse multiple correlated features for continuous bridge monitoring [11,12]. The sequential probability ratio test, based on Wald's sequential analysis [13], and sequential change-point detection methods were also proposed in succession [14,15].

However, most existing methods used for early warning rely on a single-factor indicator related to structural performance and trigger an alarm only when the monitored indicator exceeds a preset, fixed threshold [16-21]. Although the aforementioned sequential monitoring schemes can partially satisfy the manager's requirement of bridge safety by controlling the false alarm rate and reducing the detection delay, this mode is essentially reactive and exposes the structure to a higher failure risk, since damage may have already occurred by the time the alarm is triggered. Moreover, determining the threshold solely based on a fixed safety limit often neglects the quantification of the economic consequences of warning decisions, and also fails to account for the significant uncertainties arising from environmental variability, modeling limitations, measurement noise, and other sources [22-25]. The uncertainty in the structural state estimate evolves as monitoring data accumulate, and the costs induced by warning responses also change continuously. A fixed threshold cannot adapt to such time-varying uncertainty and cost, and therefore cannot maintain optimal cost-effectiveness in the long term. These limitations essentially stem from the fact that most existing methods treat early warning merely as a passive detection task. In practice, a desirable early warning scheme should shift from passive alarming to proactive decision-making. With monitoring data accumulating continuously, both the uncertainties and the costs evolve over time, and the bridge manager should therefore assess, at each decision epoch, whether an early warning should be issued against the prospective structural failure risk. Crucially, successive decisions are not independent, since each decision reshapes the state and the cost structure on which subsequent decisions are conditioned. Early warning of bridge failure risk thus constitutes a sequential decision-making problem, in which each decision is made under uncertainties and affects the costs and risks of subsequent decisions.

Sequential decision-making problems have been extensively studied in infrastructure management, where they are usually formulated as Markov decision processes (MDPs) or partially observable MDPs (POMDPs) for inspection and maintenance planning [26-29]. Despite their theoretical rigor, these formulations face several practical difficulties. The state and observation spaces must be discretized, and solving the resulting dynamic programming problem becomes computationally expensive as the problem size grows. Deep reinforcement learning has been adopted to alleviate the computational burden by approximating value functions and policies with deep neural networks [29], but it only offers limited interpretability and lacks rigorous uncertainty quantification. In addition, the transition and observation models in these formulations are generally assumed to be known and fixed, so model epistemic uncertainties are ignored and the model parameters are not updated when new monitoring data become available [30]. Furthermore, the solution of a POMDP is often a policy that is difficult for decision makers to interpret and implement in routine operation [31]. These limitations motivate the development of a sequential decision-making framework that properly addresses evolving uncertainties, iteratively refines decision models and actionable decisions with incoming data, and ensures enough computational efficiency and interpretability.

The Bayesian paradigm constitutes such a framework, delivering a rational and robust machinery to quantify and update structural state uncertainties to facilitate decision procedures. The probability assigned to each possible structural state measures its relative plausibility to other possible states for decision-making purposes, and it can be iteratively updated to facilitate decision revision as observational evidence accumulates [32-35]. Building upon this foundation, Bayesian decision theory [36,37] provides a general framework for decision-making under uncertainties. Rather than relying on a single-factor indicator, optimal decisions are identified by evaluating the expected utilities of all alternative actions over the probabilistic characterization of the structural state. More importantly, this decision-making process can be implemented iteratively over time. Decisions are derived from updated state estimates and expected utilities, and are further revised upon the arrival of new monitoring data. A closed loop of data collection, model updating, and decision-making is thus formed, in which both the structural state model and the resulting decisions evolve with the accumulating data.

Beyond above advantages, pre-posterior analysis embedded within the Bayesian decision process [38,39] enables warning decisions to leverage information from monitoring data that have not yet been acquired. Potential realizations of future structural responses are simulated, and the corresponding decision under each realization is evaluated in advance. The early warning decision can thus be made according to the predicted evolution of the structural state, which shifts early warning from reactive alarming to proactive prevention. This paradigm shift from passive detection to proactive decision-making is particularly vital for the early warning of structural failure risk.

Pre-posterior and value of information analyses have been extensively investigated in structural reliability and structural health monitoring. Representative studies have quantified the value of monitoring information using structural reliability methods [40] and Bayesian model updating [41], and have further developed frameworks for quantifying the value of vibration-based SHM [42]. Value of information analysis has also been incorporated into sequential decision-making for component inspection, permanent monitoring, and system-level scheduling [43], while adaptive inspection and maintenance planning has been developed for deteriorating structural systems [44]. These studies provide important methodological foundations for decision-making under uncertainty. In the context of early warning of bridge failure risk, however, particular attention must be given to the interaction between evolving uncertainties, prospective failure risk, and the consequences of warning decisions. Successive decisions are not independent, since each decision affects the structural operation and the cost structure on which subsequent decisions are conditioned. An effective early-warning approach therefore requires consistent uncertainty propagation across decision epochs, efficient evaluation of the expected costs of sequential decisions, and a mechanism for revising decisions as new monitoring information becomes available.

This study develops a Bayesian sequential decision-making framework for early warning of structural failure risk in time-varying structural systems. The framework integrates the probabilistic state transition model, cost model, observation prediction model, and threshold-based decision strategy into an iteratively updated closed loop. The multi-source uncertainties are characterized by a belief state, which carries the information from the monitoring and action history and is propagated through a probabilistic state transition model at each decision epoch. The expected costs of sequential warning decisions are evaluated over the future structural responses predicted by a Bayesian dynamic linear model (DLM) [45] through pre-posterior analysis, and the warning threshold is optimized to balance the failure risk against the warning costs. As monitoring data accumulate, the belief state is updated via Bayes' theorem, and the warning threshold is adaptively optimized to accommodates the evolving uncertainties and costs, so that each decision is conditioned on the latest information. The main contributions of this work are: the early warning of structural failure risk is rigorously formulated as a sequential decision-making problem under uncertainties, which shifts the warning paradigm from passive detection to proactive decision-making; a Bayesian framework is developed for the solution of this problem, which enables adaptive decision-making based on the latest information; an optimization algorithm is proposed for accelerating the computation, and a theoretical improvement is established to demonstrate that the adaptive threshold strategy performs no worse than any fixed threshold strategy.

This paper is organized as follows. Section 2 presents the Bayesian sequential decision-making framework in detail, including the cost models of early warning decisions, the probabilistic structural state transition model, the optimization of the warning threshold, and the adaptive updating of the framework. Section 3 validates the effectiveness of the proposed framework through two case studies. The first examines the early warning of fatigue fracture of steel truss member, and the second examines the early warning of stay-cable failure risk using long-term monitoring data. Section 4 concludes the paper with a summary of the main findings and future research directions.

## 2. Bayesian sequential decision-making framework for early warning of structural failure risk

### 2.1. Problem formulation

Consider an in-service bridge whose structural condition evolves over a finite service horizon. Time is discretized into a sequence of decision epochs $\tau = 1, \dots, T$, at which the bridge manager evaluates the available monitoring information and decides whether an early warning action should be issued. Because the actual deterioration state of the structure cannot be directly observed and may evolve in response to management actions, the warning problem is formulated as a partially observable sequential decision-making problem.

The proposed framework is built upon a probabilistic state transition model that describes the time-dependent deterioration of the structure under a candidate action, an observation prediction model that characterizes prospective monitoring data, an action set that defines the alternative early warning actions, a cost model that quantifies the consequences of each action, and a threshold-based decision strategy that maps the current information state to an early warning action. These elements form a closed-loop process encompassing updating, prediction, optimization, and decision execution.

Specifically, let $\mathbf{X}_\tau \in \mathcal{X}$ denote the latent structural state at decision epoch $\tau$, and $Y_\tau$ denote the observation obtained during the $\tau$-th decision interval. The information available to the decision maker at epoch $\tau$ is represented by $D_\tau = \{y_{1:\tau}, a_{1:\tau-1}\}$, where $y_{1:\tau}$ denotes the observations collected up to epoch $\tau$, and $a_{1:\tau-1}$ denotes the actions taken before epoch $\tau$. Two possible early warning actions are represented by $\mathcal{A} = \{a_n, a_w\}$, where $a_n$ denotes that the warning action is not issued and the bridge keeps open, and $a_w$ denotes issuing an early warning and implementing the closure measure. Since the structural state $\mathbf{X}_\tau$ cannot be directly observed, the decision at epoch $\tau$ is based on the belief state:

$$b_\tau(\mathbf{x}) = p(\mathbf{X}_\tau = \mathbf{x} | D_\tau, F_\tau = 0), \qquad \mathbf{X}_\tau \in \mathcal{X} \tag{1}$$

where $F_\tau = 1$ is used to indicate that the structure has failed , and $F_\tau = 0$ denotes that the bridge has survived up to epoch $\tau$. The belief state summarizes all information contained in the monitoring and action history and serves as the information state for sequential decision-making. The warning strategies $\pi = \{\pi_\tau\}_{\tau=1}^{T}$ are defined as a sequence of mappings from belief states to actions:

$$a_\tau = \pi_\tau(b_\tau) \tag{2}$$

The objective is therefore to optimize the strategies $\pi$. For engineering interpretability and computational tractability, the decision strategy is parameterized by a warning threshold. Let $R_\tau(b_\tau)$ denote a predicted risk function (introduced in Section 2.3), a threshold-based warning strategy is defined as:

$$\pi_\tau(b_\tau; s_\tau) = \begin{cases} a_n, & R_\tau(b_\tau) < s_\tau \\ a_w, & R_\tau(b_\tau) \geq s_\tau \end{cases} \tag{3}$$

where $s_\tau$ is the warning threshold at decision epoch $\tau$. The warning threshold $s_\tau$ is a decision threshold on the predicted failure risk, rather than a physical limit-state threshold that separates safe and failed structural states. It partitions the risk space into a no-warning region and a warning region, i.e., a warning is issued when $R_\tau(b_\tau) \geq s_\tau$, whereas operation continues without a warning otherwise. The threshold-based representation reduces the complexity of belief-to-action mapping, and is consistent with the warning mechanisms commonly used in bridge management. The specific physical meaning of this threshold will be discussed in Section 2.3.

Let $C_\tau(\mathbf{X}_\tau, a_\tau)$ denote the cost incurred at decision epoch $\tau$. For a threshold-based strategy $\pi = \{\pi_\tau(b_\tau; s_\tau)\}_{\tau=1}^{T}$, the expected discounted total cost is:

$$J(\pi) = \mathbb{E}\left[\sum_{\tau=1}^{T_s} \gamma^\tau C_\tau\big(\mathbf{X}_\tau, \pi_\tau(b_\tau; s_\tau)\big) \,|b_1\right] \tag{4}$$

where $\gamma$ is the discount factor; and $T_s$ is the termination decision epoch. The sequential optimization problem is therefore represented as:

$$\pi^* = \underset{\pi\in\Pi}{\operatorname{argmin}}\, J(\pi) \tag{5}$$

where $\Pi$ denotes the admissible class of threshold-based strategies. Alternatively, the sequential decision problem can be represented by the value function [46]:

$$V_\tau(b_\tau) = \min\{\bar{C}_{w,\tau}(b_\tau), \bar{C}_{n,\tau}(b_\tau) + \gamma[1 - q_\tau(b_\tau, a_n)]\mathbb{E}[V_{\tau+1}(b_{\tau+1})|b_\tau, a_n]\} \tag{6}$$

where $\bar{C}_{w,\tau}(b_\tau)$ is the expected cost of issuing a warning; $\bar{C}_{n,\tau}(b_\tau)$ is the expected cost of not trigger a warning; and $q_\tau(b_\tau, a_n)$ is the conditional failure probability during the current decision interval. The first term represents stopping operation through an early warning, whereas the second term represents continuing operation, acquiring additional information, and preserving future decision opportunities.

The overall framework for addressing such a sequential decision-making problem is illustrated in Figure 1. The framework consists of four interconnected modules: Bayesian belief updating, pre-posterior prediction, threshold optimization, and warning-decision execution. At decision epoch $\tau$, the monitoring and action history $D_\tau$ is used to update the belief state $b_\tau$, which represents the posterior distribution of the latent structural state $\mathbf{X}_\tau$. Conditioned on the updated belief state, the probabilistic state transition model and the observation prediction model generate possible future structural states and monitoring observations. For each simulated scenario, the belief state is propagated over the remaining decision horizon, and the costs associated with each decision are evaluated. The expected total cost of a candidate warning threshold is estimated by averaging the costs over these scenarios, and the optimal threshold $s_\tau^*$ is obtained by minimizing the expected total cost. The current warning action is then selected by comparing the current risk measure with optimal threshold. After the selected action is executed, newly collected monitoring data are incorporated at the next decision epoch, and the same procedure is repeated. This closed-loop implementation enables the belief state and warning threshold to be adaptively updated as new information becomes available.

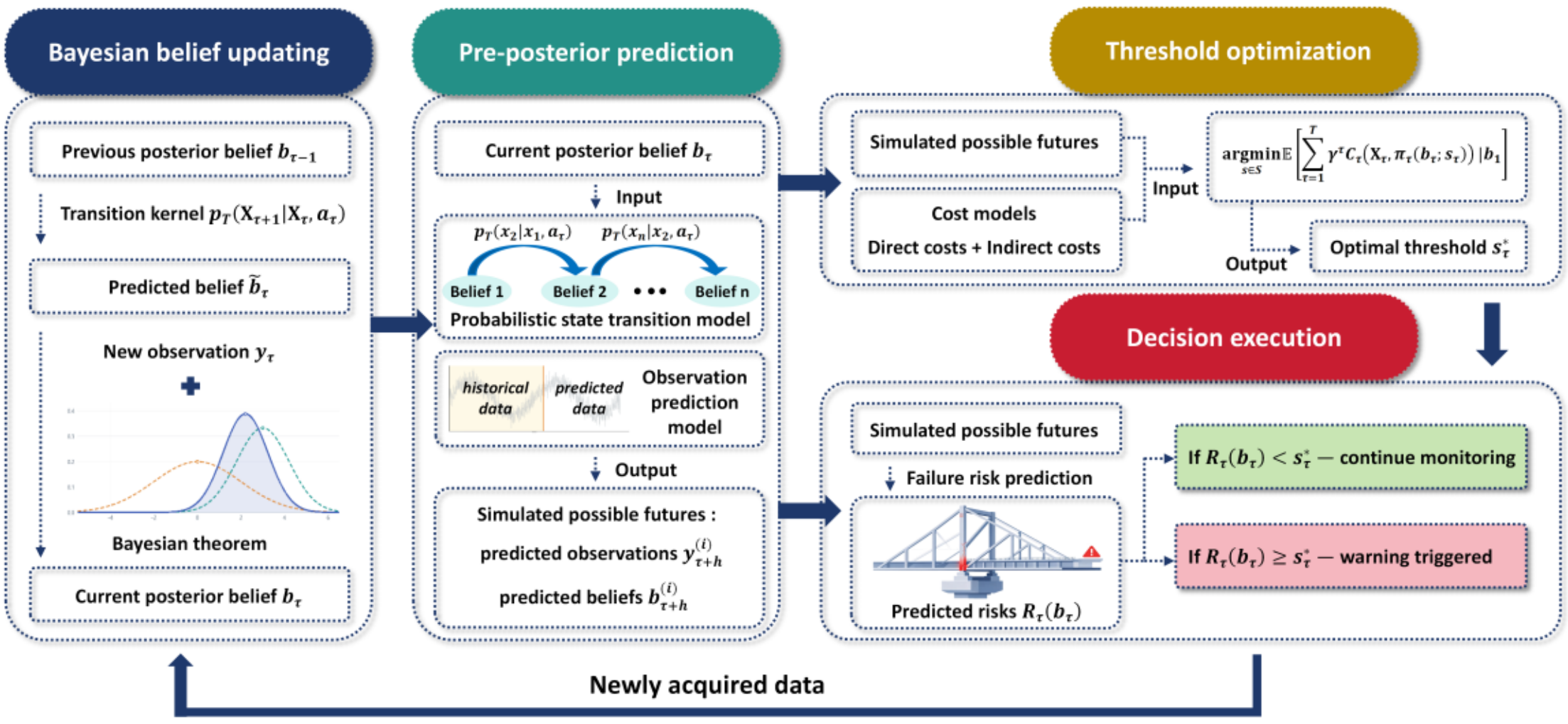

Figure 1 Overview of the proposed adaptive Bayesian sequential decision-making framework. At each decision epoch, monitoring data are used to update the belief state; future structural states and observations are predicted for pre-posterior cost evaluation; the warning threshold is optimized by minimizing the expected total cost; and the current risk is compared with the optimized threshold to select and execute a warning action. Newly acquired data then initiate the next updating cycle.

## 2.2. Basic models for early warning decision-making

### 2.2.1. Probabilistic structural state transition model and Bayesian belief updating

The sequential decision-making formulation in Section 2.1 requires a model that propagates the current belief state under a candidate action and evaluates the probability of structural failure during the subsequent decision interval. This section establishes the probabilistic state transition model, the observation model, and the corresponding Bayesian belief updating procedure. The detailed fatigue reliability analysis method used in the case studies is provided in **Appendix B**. An example of structural deterioration model.

According to the definition in Section 2.1, $\mathbf{X}_\tau \in \mathcal{X}$ denote the latent structural state at decision epoch $\tau$. For fatigue deterioration, $\mathbf{X}_\tau$ may cover, for example, the equivalent fatigue crack depth, the statistics of the equivalent stress range, material parameters, and the critical crack depth. Other deterioration mechanisms can be incorporated by redefining $\mathbf{X}_\tau$.

The structural state evolves according to a stochastic transition model $p_T(\mathbf{X}_{\tau+1}|\mathbf{X}_\tau, a_\tau)$ [26,27,47], which depends on the action at epoch $\tau$. This transition distribution is induced by the deterioration model and the random load effects during the decision interval. The action $a_\tau$ modifies the distribution by changing the traffic condition. For example, a simplified representation is $\Delta\sigma_\tau(a_\tau) = \rho(a_\tau)\Delta\sigma_\tau^{(0)}$, where $\Delta\sigma_\tau^{(0)}$ denotes the stress range under normal operation and $\rho(a_\tau)$ is a factor depend on the action. For the no warning action, $\rho(a_n)$ equals 1; and for a warning action involving bridge closure, $\rho(a_w)$ is between 0 and 1.

When the current structural state is known, the distribution of the next state is given directly by the transition kernel $p_T(\mathbf{x}'|\mathbf{x}, a_\tau)$. However, in practical applications, the current state is uncertain and is characterized by the belief state $b_\tau(\mathbf{x})$ here. The distribution of $\mathbf{X}_{\tau+1}$ conditional on the available information $D_\tau$ and the selected action $a_\tau$ can be obtained using the law of total probability:

$$p(\mathbf{X}_{\tau+1} = \mathbf{x}'|D_\tau, a_\tau) = \int_{\mathcal{X}} p(\mathbf{X}_{\tau+1} = \mathbf{x}'|\mathbf{X}_\tau = \mathbf{x}, D_\tau, a_\tau)\, p(\mathbf{X}_\tau = \mathbf{x}|D_\tau, a_\tau) d\mathbf{x} \quad (7)$$

According to the Markov property of the state-transition model, the first term in the integral reduces to $p_T(\mathbf{x}'|\mathbf{x}, a_\tau)$. In addition, the action selected at epoch $\tau$ does not change the information already contained in $D_\tau$, and therefore the predicted belief can be obtained by:

$$\tilde{b}_{\tau+1}(\mathbf{x}'|a_\tau) = \int_{\mathcal{X}} p_T(\mathbf{x}'|\mathbf{x}, a_\tau)\, b_\tau(\mathbf{x}) d\mathbf{x} \quad (8)$$

where $b_\tau(\mathbf{x})$ is the belief state given in Eq. (1), and $\tilde{b}_{\tau+1}$ is the predicted belief distribution of $\mathbf{X}_{\tau+1}$ before incorporating the monitoring observation at epoch $\tau + 1$ (i.e., the prior distribution at epoch $\tau + 1$).

After the state transition, a new monitoring observation is obtained. Let $Y_{\tau+1}$ denote the predicted observation that will be obtained after the state transition from epoch $\tau$ to $\tau + 1$. Depending on the monitoring system, $Y_{\tau+1}$ may represent strain measurements, stress range statistics, cable-force measurements, or other damage-sensitive features, and it's represented as a scalar here. The predicted observation is related to the latent structural state through the following equation:

$$Y_{\tau+1} = h_Y(\mathbf{X}_{\tau+1}, a_\tau) + o_{\tau+1} \quad (9)$$

where $h_Y(\cdot)$ denotes the observation function for predicting $Y_{\tau+1}$ and $o_{\tau+1}$ denotes the uncertain measurement noise and uncertain discrepancy because of prediction model imperfections. The specific form of the observation function and noise distribution should be selected according to the monitoring variable and data processing procedure. If the predicted observation is a statistic of an intermediate uncertain quantity, such as the uncertain stress range, the likelihood can be obtained by marginalizing that intermediate quantity.

Once the actual monitoring observation $y_{\tau+1}$ becomes available, the posterior belief is obtained through Bayes' theorem, which follows the general Bayesian model updating framework used for structural systems [48]:

$$b_{\tau+1}(\mathbf{x}') = \frac{p_O(y_{\tau+1}|\mathbf{x}', a_\tau)\tilde{b}_{\tau+1}(\mathbf{x}'|a_\tau)}{p(y_{\tau+1}|\mathcal{D}_\tau, a_\tau)} \tag{10}$$

The normalizing constant $p(y_{\tau+1}|\mathcal{D}_\tau, a_\tau)$ can be computed by:

$$p(y_{\tau+1}|\mathcal{D}_\tau, a_\tau) = \int_{\mathcal{X}} p_O(y_{\tau+1}|\mathbf{x}'', a_\tau)\tilde{b}_{\tau+1}(\mathbf{x}''|a_\tau)d\mathbf{x}'' \tag{11}$$

Therefore, the complete probabilistic update from the current belief about $\mathbf{X}_\tau$ to the posterior belief about $\mathbf{X}_{\tau+1}$ can be written as:

$$b_{\tau+1}(\mathbf{x}') = \frac{p_O(y_{\tau+1}|\mathbf{x}', a_\tau)\int_{\mathcal{X}} p_T(\mathbf{x}'|\mathbf{x}, a_\tau)\, b_\tau(\mathbf{x})d\mathbf{x}}{\int_{\mathcal{X}} p_O(y_{\tau+1}|\mathbf{x}'', a_\tau)\left[\int_{\mathcal{X}} p_T(\mathbf{x}''|\mathbf{x}, a_\tau)\, b_\tau(\mathbf{x})d\mathbf{x}\right]d\mathbf{x}''} \tag{12}$$

For notational convenience, the complete update is denoted by the operator:

$$b_{\tau+1} = B(b_\tau, a_\tau, y_{\tau+1}) \tag{13}$$

where $B(\cdot)$ represents both the state transition and the Bayesian updating process. This update procedure follows the general Bayesian filtering framework, in which the state distribution is first propagated through the transition model and subsequently corrected using the newly acquired observation [49,50].

In fatigue deterioration applications where the observation $Y_{\tau+1}$ is defined as the sample mean of equivalent stress ranges obtained from rainflow counting within the $(\tau+1)$-th monitoring period, suppose that the likelihood function depends on the structural state only through the mean equivalent stress range $\mu$, and that the stress range cycles are Gaussian with variance $\sigma_0^2$ estimated from monitoring data. Under these assumptions, the observation equation in Eq. (9) takes the specific form:

$$Y_{\tau+1} = \mu + o_{\tau+1}, \qquad o_{\tau+1} \sim \mathcal{N}\left(0, \frac{\sigma_0^2}{N_{\tau+1}}\right) \tag{14}$$

where $N_{\tau+1}$ is the number of rainflow cycles in the monitoring period. Here, $o_{\tau+1}$ represents the random deviation of the sample mean from the underlying mean $\mu$. Its zero-mean Gaussian distribution and its variance follow directly from the definition of $Y_{\tau+1}$ as the sample mean of $N_{\tau+1}$ Gaussian stress-range cycles. When the individual stress range cycles follow a non-Gaussian distribution, the same Gaussian form can be adopted as a large-sample approximation to the distribution of the sample mean based on the central limit theorem. Furthermore, the marginal belief of $\mu$ is assumed to satisfy $\mu|\mathcal{D}_\tau \sim \mathcal{N}(\eta_\tau, \varepsilon_\tau^2)$. The posterior distribution can then be derived in a closed-form. Specifically, let $\bar{y}_{\tau+1}$ denote the sample mean of $N_{\tau+1}$ stress range cycles, then the posterior distribution satisfies $\mu|\mathcal{D}_{\tau+1} \sim \mathcal{N}(\eta_{\tau+1}, \varepsilon_{\tau+1}^2)$. The mean value and variance can be derived using standard conjugate Bayesian updating [36]:

$$\eta_{\tau+1} = \frac{\sigma_0^2 \eta_\tau + N_{\tau+1} \varepsilon_\tau^2 \bar{y}_{\tau+1}}{\sigma_0^2 + N_{\tau+1} \varepsilon_\tau^2} \tag{15}$$

$$\varepsilon_{\tau+1}^2 = \frac{\sigma_0^2 \varepsilon_\tau^2}{\sigma_0^2 + N_{\tau+1} \varepsilon_\tau^2} \tag{16}$$

Note that only the marginal belief of $\mu$ is required to be Gaussian, while the joint belief may take an arbitrary form. In this way, the load-related uncertainty is updated analytically at each decision epoch, and the remaining state uncertainty, such as that of the damage state, is propagated through the transition model and indirectly refined by the updated belief of $\mu$, thereby avoiding repeated numerical integration in the sequential updating process.

### 2.2.2. Observation prediction model

The value function in Section 2.1 contains the expectation of the future value function over the next monitoring observation, i.e., $\mathbb{E}[V_{\tau+1}(b_{\tau+1})|b_\tau, a_n]$. Evaluating this expectation requires characterizing the distribution of monitoring data before they are actually collected, which is referred to as pre-posterior analysis. This section establishes a prediction model based on a Bayesian dynamic linear model (DLM).

The monitoring system collects data at a finer time resolution than the decision epochs. Let $t$ index the sampling instants of the monitoring system, and $\mathbf{y}_t$ denote the raw observations collected at sampling instant $t$. The monitoring series is modeled by a Bayesian DLM, consisting of an observation equation and a state evolution equation:

$$\mathbf{y}_t = \mathbf{F}_t \mathbf{z}_t + \mathbf{v}_t, \qquad \mathbf{v}_t \sim \mathcal{N}(0, \mathbf{V}_t) \tag{17}$$

$$\mathbf{z}_t = \mathbf{G}_t \mathbf{z}_{t-1} + \mathbf{w}_t, \qquad \mathbf{w}_t \sim \mathcal{N}(0, \mathbf{W}_t) \tag{18}$$

where $\mathbf{z}_t$ denotes the latent state of the observation series, which represents its statistical dynamics, such as the local level, trend, and periodic components induced by traffic and environmental conditions; $\mathbf{F}_t$ is the observation matrix; and $\mathbf{G}_t$ is the state evolution matrix. The noise $\mathbf{v}_t$ and $\mathbf{w}_t$ are modeled as independent zero-mean Gaussian vectors with covariance matrices $\mathbf{V}_t$ and $\mathbf{W}_t$, respectively. This joint noise probability model gives the largest uncertainty (maximizes Shannon's information entropy) for $\mathbf{v}_t$ subject to the first two moment constraints: $\mathbb{E}[\mathbf{v}_t] = 0$ and $\mathbb{E}[\mathbf{v}_t \mathbf{v}_t^{\mathrm{T}}] = \mathbf{V}_t$ ( and likewise for $\mathbf{w}_t$). It thus introduces the minimal additional assumptions beyond the available moment information, avoiding unwarranted specification of the error distribution. In addition to $\mathbf{v}_t$ and $\mathbf{w}_t$ being mutually independent, they are modelled as independent of the initial state $\mathbf{z}_0 \sim \mathcal{N}(\mathbf{m}_0, \Sigma_0)$. Under the linearity and Gaussian assumptions, the posterior distribution of the state given the monitoring series up to the last monitoring instance during decision epoch $\tau$ remains Gaussian, i.e., $p(\mathbf{z}_\tau | \mathcal{D}_\tau) = \mathcal{N}(\mathbf{m}_\tau, \Sigma_\tau)$, and its hyper-parameters for the covariance matrices $\mathbf{V}_t$ and $\mathbf{W}_t$ are updated sequentially via Bayes' theorem through the Kalman filter recursions [49,50]. The matrices $\mathbf{F}_t$ and $\mathbf{G}_t$ are specified according to the characteristics of the monitoring variable, and the unknown hyper-parameters are estimated from the historical monitoring data.

It should be emphasized that $\mathbf{z}_t$ is a statistical state of the monitoring data series, which is distinct from the physical structural state $\mathbf{X}_\tau$ defined in Section 2.1. The former serves to generate prospective monitoring observations for pre-posterior analysis, whereas the latter characterizes the deterioration condition. The observation $Y_\tau$ defined in Section 2.1 is a statistic or feature computed from the raw observations collected within the corresponding decision interval:

$$Y_\tau = \mathcal{G}(\{\mathbf{y}_t : t \in (\tau - 1, \tau]\}) \tag{19}$$

where $\mathcal{G}(\cdot)$ denotes the data processing operator, which extract a statistic or feature from the raw observations.

To evaluate the future value function and support the pre-posterior analysis, the monitoring observations over the subsequent decision intervals need to be predicted. Given the historical monitoring series up to decision epoch $\tau$, the predicted observation satisfies $\mathbf{y}_t|\mathbf{y}_{1:\tau}\sim\mathcal{N}(\mathbf{f}_t,\mathbf{Q}_t)$, where the predictive mean $\mathbf{f}_t$ and covariance matrix $\mathbf{Q}_t$ are obtained by propagating the filtered state $\mathcal{N}(\mathbf{m}_\tau,\sum_\tau)$ forward through the system equation. By sampling from the predicted observation distribution, a sample of observation data can be obtained using the data processing operator:

$$y_{\tau+h}^{(i)}=\mathcal{G}\left(\left\{\mathbf{y}_t^{(i)}:t\in(\tau+h-1,\tau+h]\right\}\right),\qquad h=1,\dots,H \tag{20}$$

where $H$ denotes the number of future decision intervals. The resulting samples $y_{\tau+h}^{(i)}, i=1,\dots,M$ characterize the predictive distribution of the observations, where $M$ is the number of samples. With such a distribution, the expectation of the future value function in the Bellman equation is evaluated by averaging over the prospective monitoring data:

$$\mathbb{E}_{Y_{\tau+1}|b_\tau,a_n}\left[V_{\tau+1}\left(B(b_\tau,a_\tau,Y_{\tau+1})\right)\right]\approx\frac{1}{M}\sum_{i=1}^{M}V_{\tau+1}\left(B\left(b_\tau,a_\tau,y_{\tau+1}^{(i)}\right)\right) \tag{21}$$

where $B(\cdot)$ is the belief updating operator defined in Eq. (13). This is a pre-posterior evaluation in the sense that the hypothetical posterior belief is computed for each possible future observation, and the expectation is taken before the observation is actually collected. For multi-step scenarios, the same procedure is applied recursively along each simulated observation path $y_{\tau+1:\tau+H}^{(i)}$.

### 2.2.3. Cost model

The Bellman equation (3) in Section 2.1 requires the expected costs of the two candidate actions at each decision epoch. This section defines the cost components of the early warning problem and derives the expected costs as functions of the belief state.

An example of cost model associated with bridge failure is given in **Appendix A**. An example of cost models for bridge early warning. According to the contents in this appendix, the consequences can be classified into four cost categories: reconstruction cost $C_{RB}$, life loss cost $C_{LL}$, running cost $C_{RN}$ and time loss cost $C_{TL}$. Each decision must be evaluated under both bridge failure and non-failure scenarios. When the warning is not issued: if the bridge fails, the cost should include $C_{RB}$, $C_{LL}$, $C_{RN}$, and $C_{TL}$; if the bridge does not fail, there is no cost incurred. When an early warning is triggered: if the bridge fails, the cost comprises $C_{RB}$, $C_{RN}$, and $C_{TL}$; if the bridge does not fail, the cost should account for $C_{RN}$ and $C_{TL}$. Since the structural state is uncertain, the cost of each action is evaluated as an expectation with respect to the belief state.

For the action $a_n$ to not issue a warning, the expected cost can be computed by:

$$\bar{C}_{n,\tau}(b_\tau) = \int_{\mathcal{X}} q_\tau(\mathbf{x}, a_n)\, C_F b_\tau(\mathbf{x}) d\mathbf{x} = q_\tau(b_\tau, a_n) C_F \tag{22}$$

$$C_F = C_{RB} + C_{LL} + C_{RN} + C_{TL} \tag{23}$$

$$q_\tau(b_\tau, a_n) = \int_{\mathcal{X}} q_\tau(\mathbf{x}, a_n)\, b_\tau(\mathbf{x}) d\mathbf{x} \tag{24}$$

where $C_F$ denotes the cost when the warning is not issued and the bridge fails; $q_\tau(\mathbf{x}, a_n)$ denotes the conditional failure probability during the current decision interval given the structural state; and $q_\tau(b_\tau, a_n)$ is its expectation under the belief state, whose evaluation is detailed in the next section.

For the warning action $a_w$, the cost can be computed by:

$$\bar{C}_{w,\tau}(b_\tau) = C_{RN} + C_{TL} + q_\tau(b_\tau, a_w)(C_F - C_{LL}) \tag{25}$$

where $q_\tau(b_\tau, a_w)$ denotes the failure probability under the warning action. Since the closure measure removes the traffic loading, no additional fatigue damage is accumulated during the closure, and the failure probability $q_\tau(b_\tau, a_w)$ is often negligible. This is the also the economic basis of the threshold-based warning strategy. No warning action is preferable when the risk is low, while warning becomes preferable once the expected failure loss exceeds the warning cost.

Together with the discount factor, the expected costs $\bar{C}_{n,\tau}(b_\tau)$ and $\bar{C}_{w,\tau}(b_\tau)$ constitute the cost terms of the Bellman equation, through which the economic consequences of candidate actions are balanced against the value of future decision opportunities.

### 2.3. Solutions of the Bayesian sequential decision-making problem

#### 2.3.1. Failure risk prediction

The Bellman equation and the cost model require the conditional failure probability during each decision interval, and the threshold-based warning strategy requires a risk measure that characterizes the failure risk under the current belief state. This section establishes the failure risk prediction model based on the belief state and the state transition model.

Structural failure is defined through a limit-state function, and it occurs when the structural state enters the failure domain:

$$\mathcal{X}_f = \{\mathbf{x} \in \mathcal{X}: g(\mathbf{x}) \leq 0\} \tag{26}$$

The complementary set $\mathcal{X}_s = \mathcal{X} \backslash \mathcal{X}_f$ denotes the safe domain. For fatigue deterioration, the limit-state function is defined by the critical and current crack depths, and $g(\mathbf{x}) \leq 0$ indicates that the fatigue crack has reached its critical size. More details are provided in **Appendix B**. An example of structural deterioration model.

The failure probability during the decision interval $(\tau, \tau + 1]$ conditioned on survival up to epoch $\tau$ is obtained by integrating the predicted belief over the failure domain:

$$q_\tau(b_\tau, a_\tau) = P\big(\mathbf{X}_{\tau+1} \in \mathcal{X}_f | a_\tau, D_\tau, F_\tau = 0\big) = \int_{\mathcal{X}_f} \tilde{b}_{\tau+1}(\mathbf{x}'|a_\tau) d\mathbf{x}' \tag{27}$$

where $\tilde{b}_{\tau+1}(\mathbf{x}'|a_\tau)$ is the predicted belief obtained from the state transition model in Eq. (8). Equivalently, by interchanging the order of integration, the following equations can be obtained:

$$q_\tau(b_\tau, a_\tau) = \int_{\mathcal{X}} q_\tau(\mathbf{x}, a_\tau)\, b_\tau(\mathbf{x}) d\mathbf{x} \tag{28}$$

$$q_\tau(\mathbf{x}, a_\tau) = \int_{\mathcal{X}_f} p_T(\mathbf{x}'|\mathbf{x}, a_\tau)\, d\mathbf{x}' \tag{29}$$

where $q_\tau(\mathbf{x}, a_\tau)$ denotes the conditional failure probability given a known structural state, which is evaluated by the reliability analysis.

Since the belief state is defined conditioned on structural survival, the predicted belief must be restricted to the safe domain before being propagated over multiple intervals. Specifically, after the one-step propagation, the probability mass in the failure domain corresponds to the failure event, and the belief conditioned on survival during the interval is obtained by truncation and renormalization:

$$\tilde{b}_{S,\tau+1}(\mathbf{x}'|a_\tau) = \begin{cases} \dfrac{\tilde{b}_{\tau+1}(\mathbf{x}'|a_\tau)}{1 - q_\tau(b_\tau, a_\tau)}, & \mathbf{x}' \in \mathcal{X}_S \\ 0, & \mathbf{x}' \in \mathcal{X}_f \end{cases} \tag{30}$$

This belief ensures that the multi-step risk prediction is consistently conditioned on the structure having survived all preceding intervals, which is consistent with the definition of the belief state.

The failure risk over multiple future decision intervals is evaluated by recursively applying the state propagation and the renormalization. Starting from $b_S^{(0)} = b_\tau$, for $h = 1, \dots, H$, we have:

$$\tilde{b}^{(h)}(\mathbf{x}') = \int_{\mathcal{X}_S} p_T(\mathbf{x}'|\mathbf{x}, a) b_S^{(h-1)}(\mathbf{x}) d\mathbf{x} \tag{31}$$

$$b_S^{(h)}(\mathbf{x}') = \frac{\tilde{b}^{(h)}(\mathbf{x}')}{1 - q_{\tau+h}}, \qquad \mathbf{x}' \in \mathcal{X}_S \tag{32}$$

$$q_{\tau+h} = \int_{\mathcal{X}_f} \tilde{b}^{(h)}(\mathbf{x}') d\mathbf{x}' \tag{33}$$

The survival probability over the subsequent $H^*$ decision intervals is the product of the interval survival probabilities:

$$S_{\tau,H^*}(b_\tau) = \prod_{h=1}^{H^*} (1 - q_{\tau+h}) \tag{34}$$

The corresponding cumulative failure probability is

$$P_{F,\tau,H^*}(b_\tau) = 1 - S_{\tau,H^*}(b_\tau) \tag{35}$$

The risk measure is thus defined as the cumulative failure probability over a prescribed horizon $H^*$, i.e., $R_\tau(b_\tau) = P_{F,\tau,H^*}(b_\tau)$. The horizon $H^*$ can be selected as the remaining service life or a prescribed management period.

The multi-step risk is typically evaluated under the no warning action, since it serves to characterize the risk evolution if no warning is issued. In numerical implementation, the above recursion can be carried out by Monte Carlo simulation. When the state is propagated through the transition model, samples falling into the failure domain are removed at each step, and the failure probabilities are estimated by the corresponding sample frequencies. With the predicted failure risk $R_\tau(b_\tau)$, a proactive warning becomes possible, which is significant for ensuring structural safety and resource allocation.

### 2.3.2. Threshold optimization and adaptive implementation

The threshold-based warning strategy introduced in Eq. (3) of Section 2.1 maps the risk measure $R_\tau(b_\tau)$ to the warning action through the threshold $s_\tau$, and the sequential optimization problem seeks the threshold sequence that minimizes the expected total discounted cost within the admissible class of threshold-based strategies. This section presents the solution method for the threshold optimization problem and the corresponding adaptive implementation procedure.

Since the value function in the Bellman equation has no analytical representation, it is intractable to derive a closed-form solution. A scenario-based solution method is therefore adopted, in which the expected total cost of a candidate threshold sequence is evaluated by Monte Carlo simulation, and the optimal threshold sequence is searched over a discretized threshold space. Specifically, at decision epoch $\tau$, the expected total cost over the prescribed horizon is evaluated for a candidate warning threshold sequence $\{s_\tau\}_{\tau=0}^{T}$. Here, each threshold in the sequence is treated as an identical constant $s$. While this simplification may lead to deviations from the optimal result, it is very suitable for our framework. This will be discussed later.

The evaluation starts from the current belief state $b_\tau$ and proceeds by simulating the closed-loop decision process over $N_M$ independent scenarios. In scenario $i$, an initial state is sampled from $b_\tau$, and the following steps are repeated for each decision epoch: (1) compute the risk $R_\tau\left(b_\tau^{(i)}\right)$ by the failure risk prediction model in Section 2.3.1; (2) determine the action by the threshold strategy, i.e., issue a warning if $R_\tau\left(b_\tau^{(i)}\right) \geq s$, and continue operation otherwise; (3) if a warning is issued, add the discounted warning cost $\gamma_\tau \bar{C}_{w,\tau}$ to the cumulative cost, and terminate the scenario; (4) if operation is continued, generate the failure event with the conditional failure probability $q_\tau\left(\mathbf{x}_\tau^{(i)}, a_n\right)$, and if failure occurs, add the discounted failure cost $\gamma_\tau C_F$ and terminate the scenario; (5) otherwise, propagate the structural state through the transition model, generate the monitoring observation by the prediction model, and update the belief state by $b_{\tau+1}^{(i)} = B\left(b_\tau^{(i)}, a_n, y_{\tau+1}^{(i)}\right)$. After the simulation, the average value over all possible scenarios can be approximated by:

$$\hat{J}_\tau(s|b_\tau) = \frac{1}{N_M}\sum_{i=1}^{N_M} J_\tau^{(i)} \tag{36}$$

where $J_\tau^{(i)}$ denotes the cumulative discounted cost of scenario $i$. The estimation error is mainly controlled by the number of scenarios. Since the risk measure is a probability, the admissible threshold range is naturally bounded within $[0,1]$. At decision epoch, the optimized threshold is obtained by minimizing the estimated expected total cost over the admissible range:

$$s_\tau^* = \underset{s\in[0,1]}{\operatorname{argmin}} \hat{J}_\tau(s|b_\tau) \tag{37}$$

where $\hat{J}(s|b_\tau)$ is evaluated conditioned on the current belief state and the remaining prescribed horizon. The optimized warning threshold has both a probabilistic interpretation and a physical meaning. As stated in section 2.3.1, $R_\tau(b_\tau)$ is the cumulative failure probability over the prescribed horizon, so the warning threshold $s_\tau$ is also expressed on the same probability scale. The optimized threshold $s_\tau^*$ is interpreted as the highest predicted cumulative failure probability that can be tolerated. If $R_\tau(b_\tau) < s_\tau^*$, the predicted risk remains within the range acceptable for continued operation, and no warning is issued. Once $R_\tau(b_\tau)$ reaches or exceeds $s_\tau^*$, the predicted risk is no longer considered acceptable for continued operation, and a warning is triggered. The optimized threshold therefore defines the decision boundary between keeping operation and triggering a warning, any deviation from $s_\tau^*$ that changes the resulting warning decisions would move the decision results away from the cost-minimizing solution and lead to a higher expected total cost.

The scenario-based evaluation can be regarded as an implementation of pre-posterior analysis, since the prospective monitoring data is predicted and the cost of each possible scenario is averaged. In practical applications, the framework is implemented in a rolling manner. At each decision, the bridge manager collects the monitoring data within the current interval and updates the belief state. The threshold is also optimized based on the updated belief state and the remaining horizon. The process repeats upon the arrival of new monitoring data. In this way, the warning decision at each epoch is based on the latest information, and the threshold is adaptively optimized, thereby forming the closed loop of update, prediction, optimization, and decision execution. This adaptive implementation can be interpreted in the context of rollout and policy improvement in dynamic programming. Rollout provides a standard approach to constructing an improved policy relative to a base policy, and its cost-improvement property has been established in the dynamic programming literature [51]. In the present framework, a fixed threshold policy can be regarded as the base policy, while threshold re-optimization using the updated belief state yields an adaptive threshold policy. We formalize the corresponding cost-improvement property under explicit assumptions on the decision horizon, policy evaluation, model premises, candidate threshold set, and Bayesian belief updating in **Appendix C**. . The proof and the scope of this property are also provided in **Appendix C**. .

However, some details should be discussed. Restricting the optimization to threshold-based strategies constrains the strategy space. The threshold is optimized only within the threshold strategy class. A global optimality over all strategies would require additional conditions, which are not pursued in this study. The threshold-based strategy form is retained because of its engineering interpretability and its consistency with the warning mechanisms used in bridge management practice. In addition, although treating the threshold as a constant over the remaining horizon does not attain the full-horizon optimality of a threshold sequence, it substantially reduces the computational cost. Moreover, owing to the rolling scheme adopted in the proposed framework, the threshold can be re-optimized using the latest information at each decision epoch, so that the warning strategy adapts to the evolving uncertainty through adaptive threshold update.

### 2.3.3. Efficient threshold optimization via a shared-scenario algorithm

A direct implementation of the algorithm in section 2.3.2 repeats the closed-loop scenario simulation for every candidate threshold in the threshold set, so that the computational cost grows linearly with the number of candidate thresholds. This repetition can be avoided by exploiting a structural property of the warning decision, and an efficient optimization procedure is thus developed.

In the scenario-based evaluation, the risk prediction is computed under the no-warning action, and a warning decision terminates the scenario. Consequently, the evolution of a scenario, including the propagated structural state, the generated observations, the updated belief states, the resulting risk sequence, and the failure event, is entirely independent of the candidate threshold. The threshold only determines the epoch at which the scenario is terminated. Building on this property, a shared-scenario threshold optimization approach is developed, in which a single ensemble of scenarios serves the evaluation of all candidate thresholds. Specifically, at decision epoch $\tau$, a total of $N_M$ independent scenarios are simulated over the remaining horizon, and two quantities are recorded for each scenario. The first is the risk sequence $\left\{R_k^{(i)}\right\}$ of scenario $i$, where $R_k^{(i)}$ denotes the failure risk computed by the failure risk prediction model at each simulated epoch $k = \tau, \tau+1, \ldots, \tau+H$. Here $H$ is the number of future decision intervals, the value of $H$ can be adopted depending on the management period of interest and the available forecasting capability. In this study, $H$ is selected to cover the period that is most relevant to the current warning decision while maintaining reasonable reliability of the predicted monitoring observations. This choice balances the inclusion of future risk and cost consequences against the increase in predictive uncertainty and computational demand associated with a longer evaluation horizon.

Since the belief state is updated at each simulated epoch, this sequence characterizes how the predicted failure risk evolves. The second is the failure epoch $\tau_f^{(i)}$ of scenario $i$, defined as the first epoch at which the failure event occurs, where the failure event is sampled at each continued epoch with the conditional failure probability. If no failure occurs within the horizon, $\tau_f^{(i)} = \infty$. These two quantities are sufficient to reconstruct the cost of scenario $i$ under any candidate threshold. Under a candidate threshold $s$, a warning is issued at the first epoch at which the risk exceeds the threshold, i.e.,

$$\tau_w^{(i)}(s) = \min\left\{k \in \{\tau, \tau+1, \ldots, \tau+H\} \mid R_k^{(i)} \geq s\right\} \tag{38}$$

The scenario then terminates at epoch $\min\left\{\tau_w^{(i)}(s), \tau_f^{(i)}\right\}$, and the cumulative discounted cost of scenario $i$ under threshold $s$ is given by:

$$J^{(i)}(s) = \begin{cases} \gamma^{\tau_w^{(i)}(s)-\tau} \bar{C}_{w,\tau_w^{(i)}}\left(b_{\tau_w^{(i)}}\right), & \tau_w^{(i)}(s) \leq \tau_f^{(i)} \\ \gamma^{\tau_f^{(i)}-\tau} C_F, & \tau_f^{(i)} < \tau_w^{(i)}(s) \\ 0, & otherwise \end{cases} \tag{39}$$

The warning is taken to precede failure when the two epochs coincide, which is consistent with the order of the decision steps in the scenario simulation. All candidate thresholds are thus evaluated from the same ensemble of scenarios, and the estimated expected total cost can be computed by the following equation without repeating the simulation for each candidate value:

$$\hat{J}_\tau(s|b_\tau) = \frac{1}{M}\sum_{i=1}^{M} J^{(i)}(s) \tag{40}$$

Moreover, the cost $J^{(i)}(s)$ of each scenario is piecewise constant in $s$, with breakpoints at the failure risk values $\left\{R_k^{(i)}\right\}$. As $s$ decreases below $\left\{R_k^{(i)}\right\}$, the warning epoch of scenario $i$ shifts to an earlier epoch, while between two consecutive breakpoints the warning epoch and the cost remain unchanged. The estimated expected total cost $\hat{J}(s \mid b_\tau)$ is therefore a piecewise constant function of $s$ with at most $N_M \times (H+1)$ breakpoints. According to this structure, the minimizer of the estimated expected total cost over the admissible range is attained at one of the breakpoint values, and the optimized threshold is obtained exactly by sorting the breakpoints and scanning the resulting segments once. The overall computational cost of the threshold optimization at each decision epoch is changed from $O\big(C_R \cdot |\mathcal{S}| \cdot N_M \cdot (H+1)\big)$ for the direct implementation to $O\Big(C_R \cdot N_M \cdot (H+1) + N_M \cdot (H+1) \cdot \log\big(N_M \cdot (H+1)\big)\Big)$, where $C_R$ denote the computational cost of one nested risk prediction $R_k^{(i)}$, and $|\mathcal{S}|$ denotes the number of candidate thresholds in the threshold set. Generally, the computational cost of the shared-scenario algorithm is substantially lower than that of the direct implementation when the number of candidate thresholds is large or the nested risk prediction is computationally expensive.

This shared-scenario algorithm exactly minimizes the empirical piecewise-constant cost function for a given scenario ensemble, rather than the exact expected cost. No grid discretization error is introduced, and a discretized threshold set is no longer required because the breakpoints are generated directly by the simulation. The use of shared scenarios also ensures that comparisons among candidate thresholds within a single optimization are not affected by independent sampling across thresholds. Nevertheless, the optimized empirical cost remains subject to Monte Carlo error and may be biased low because the minimum is selected from noisy cost estimates.

The calculation process of the entire method is presented in Algorithm 1.

**Algorithm 1**

Bayesian sequential decision-making for adaptive early warning of structural failure risk

**Input:** state transition model $\{p_T\}$; The DLM model; cost models $\{\bar{C}_{w,\tau}, C_F\}$; risk horizon $H^*$; service horizon $T$; number of scenarios $N_M$;discount factor $\gamma$

**Output:** threshold sequence $\{s_\tau^*\}_{\tau=1}^{T}$ and warning decisions

**Initialize** the prior belief state $b_0$ and set $\tau = 1$

**While** $\tau \leq T$ do

  Compute observation $y_\tau$ using the monitoring data within the current interval;

  Update the belief $b_\tau$ using Eq. (13);

  Update the DLM with the newly collected data;

  **for** $i = 1, \dots, N_M$ **do**

    Sample the initial state $\mathbf{x}_\tau^{(i)}$ from $b_\tau$;

    Set $\tau_f^{(i)} = \infty$;

    **for** $k = \tau, \tau + 1, \dots, \tau + H^*$ **do**

      Compute the risk $R_k^{(i)}$ using Eqs. (31)~(35);

      Generate failure event with probability $q_k^{(i)}\left(\mathbf{x}_k^{(i)}, a_k^{(i)}\right)$ using Eq. (29);

      **if** failure occurs **then**

        Set $\tau_f^{(i)} = k$; **break**

      **else**

        Propagate the state via the transition model using Eq.(8);

        Generate the observation $y_{k+1}^{(i)}$ via the DLM;

        Update the belief $b_{k+1}^{(i)}$ using Eq. (13);

      **end if**

    **end for**

  **end for**

  Collect all breakpoints $\left\{R_k^{(i)}\right\}$ and sort them in ascending order;

  Scan the sorted breakpoints once and compute the cost using Eq. (38) and Eq. (39)

  Obtain the optimal threshold $s_\tau^*$ using Eq. (37);

  Compute the current risk $R_\tau(b_\tau)$ using Eqs. (31)~(35);

  **if** $R_\tau(b_\tau) \geq s_\tau^*$ **then**

    Issue the early warning and implement closure; **break**

  **else**

    keep the bridge open;

    $\tau = \tau + 1$

  **end if**

**end while**

**Return** $\{s_\tau^*\}_{\tau=1}^{T}$

# 3. Case studies

## 3.1 Early warning of failure risk for steel truss members of a bridge in Sweden

To validate the proposed framework, this study employs a steel riveted bascule bridge in Sweden as a representative case. The bridge has an overall length of about 68 m, with a truss main span length of 42 m and a maximum height of 7.6 m. The width of the bridge is 5.35 m. The structural system comprises two truss walls, a bridge deck consisting of longitudinal stringers connected to transverse floor beams, and a lateral bracing component. A monitoring system has been installed on this bridge [52]. The instrumentation involves 16 uniaxial strain gauges (SG01–SG16), 5 uniaxial accelerometers (A1–A5), and one inclinometer, as illustrated in Figure 2. The strain gauges are located at critical structural members to measure nominal strains at positions away from stress concentrations: SG01–SG02 and SG03–SG04 are placed on the crossbeams, SG05–SG06 are placed on the tension diagonal, SG07–SG08 are placed on the upper chord, and SG09–SG16 are located on the bridge counterweight. The accelerometers are placed at various locations to capture global and local vibration responses. The inclinometer is placed at the bridge support to record the rotation during the lifting of the bascule truss. The bridge is continuously monitored and raw data are stored in terms of 10-minute events by an acquisition system. Data from multiple bridge opening events were recorded by the monitoring system, including those collected before and after a damage event, the location of which is shown in Figure 3. The damage was found during the 58th recorded opening event and was verified as a partial cracking of a truss member in the counterweight part, where the bottom flange and a part of the web were affected. The dataset is publicly available [53]. For more details about this bridge, readers are referred to [52]. In this study, the processed stress data obtained from the strain gauges (SG01–SG16) are used. The monitoring contains records from before, during, and after the verified fracture event, which offers a way to observe the evolution of structural states.

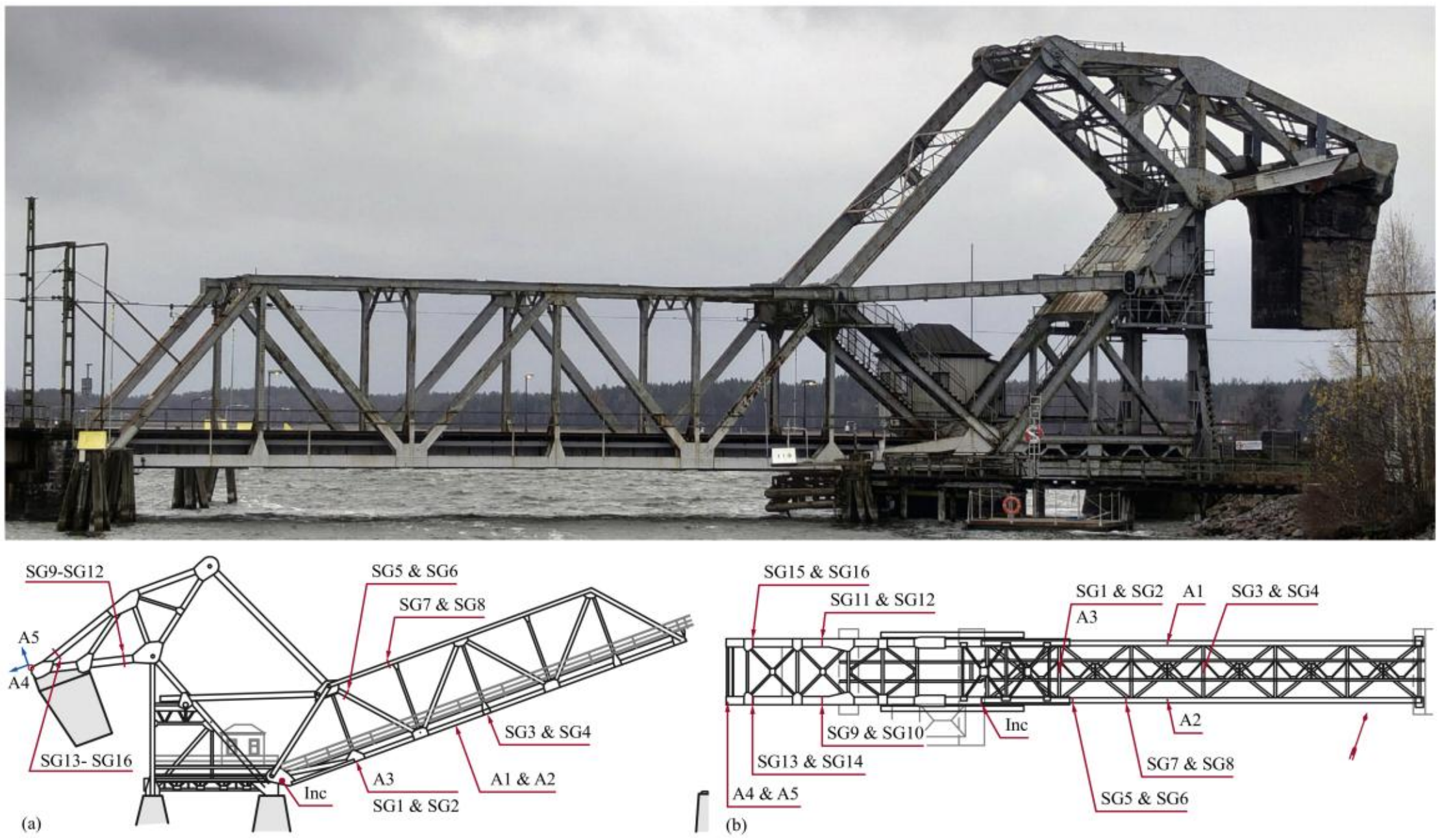

Figure 2. The overall view of the bridge along with a partial sensor layout (adapted from [52] and [53]).

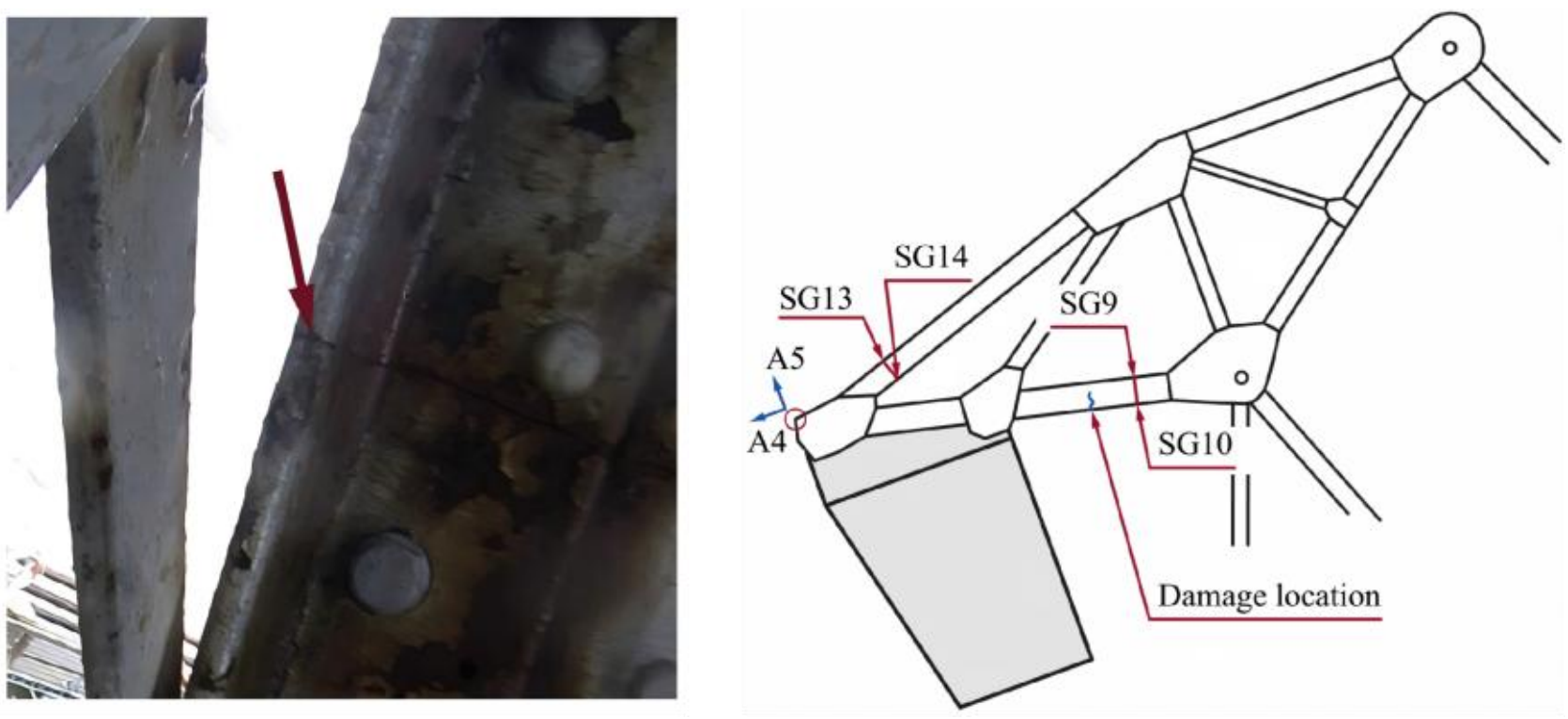

Figure 3. The specific location of the damage on the bridge. (adapted from [53])

### 3.1.1 Baseline model establishment

Following the methodology presented in Section 2, this subsection establishes the prior baseline model for the bridge using the monitoring data recorded at the first 58 opening events. The strain time histories were recorded by the sixteen strain gauges and some examples of the recorded data in the period of several bridge opening events are shown in Figure 4. Due to anomalies in the data recorded by sensors SG11 and SG16, these sensors are excluded from the presentation. Before reliability analysis, the data are processed by rainflow counting, and small cycles induced by noise are filtered out. The results for several representative samples are presented in Figure 5.

The fatigue crack growth is modeled by the Paris–Erdogan law given in **Appendix B**. An example of structural deterioration model, and the material constants are assigned according to typical values for riveted steel bridge members reported in the literature [54]. The probability models for the uncertain initial crack depth $a_0$ and the critical threshold $a_{th}$ that constitute the initial belief state $b_1$ are presented in Table 1, along with the probability models for the parameters $m$ and $C_m$. The time axis is divided into discrete decision epochs: the time point where an event ends is considered the beginning of a decision epoch.

To evaluate the economic consequences of warning decisions, the cost model in **Appendix A**. An example of cost models for bridge early warning is used. For steel railway bridges, the cost parameters are chosen to reflect railway traffic characteristics. Table 2 shows the detailed value of each parameter. $T$ represents the proportion of freight trains. The unit costs $C_3$ and $C_4$ correspond to the interruption costs per kilometer for passenger and freight trains, respectively. $O$ is the average number of passengers per train. For pure passenger trains, $C_4$, $C_6$, and $T$ are set to zero; for pure freight trains, $C_3$, $C_5$, and $O$ are set to zero. The detour length $D_L$ is typically much longer than highway detours due to the sparse railway network. With the cost parameters fixed, the expected total cost associated with each candidate threshold can be evaluated.

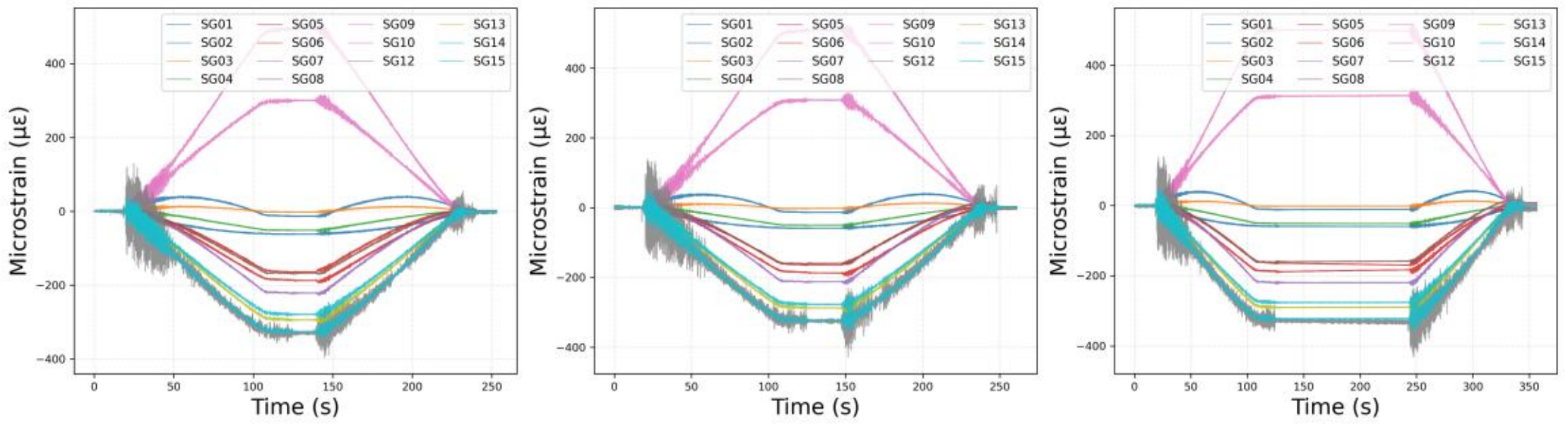

Figure 4. Examples of the recorded data in the period of several bridge open events.

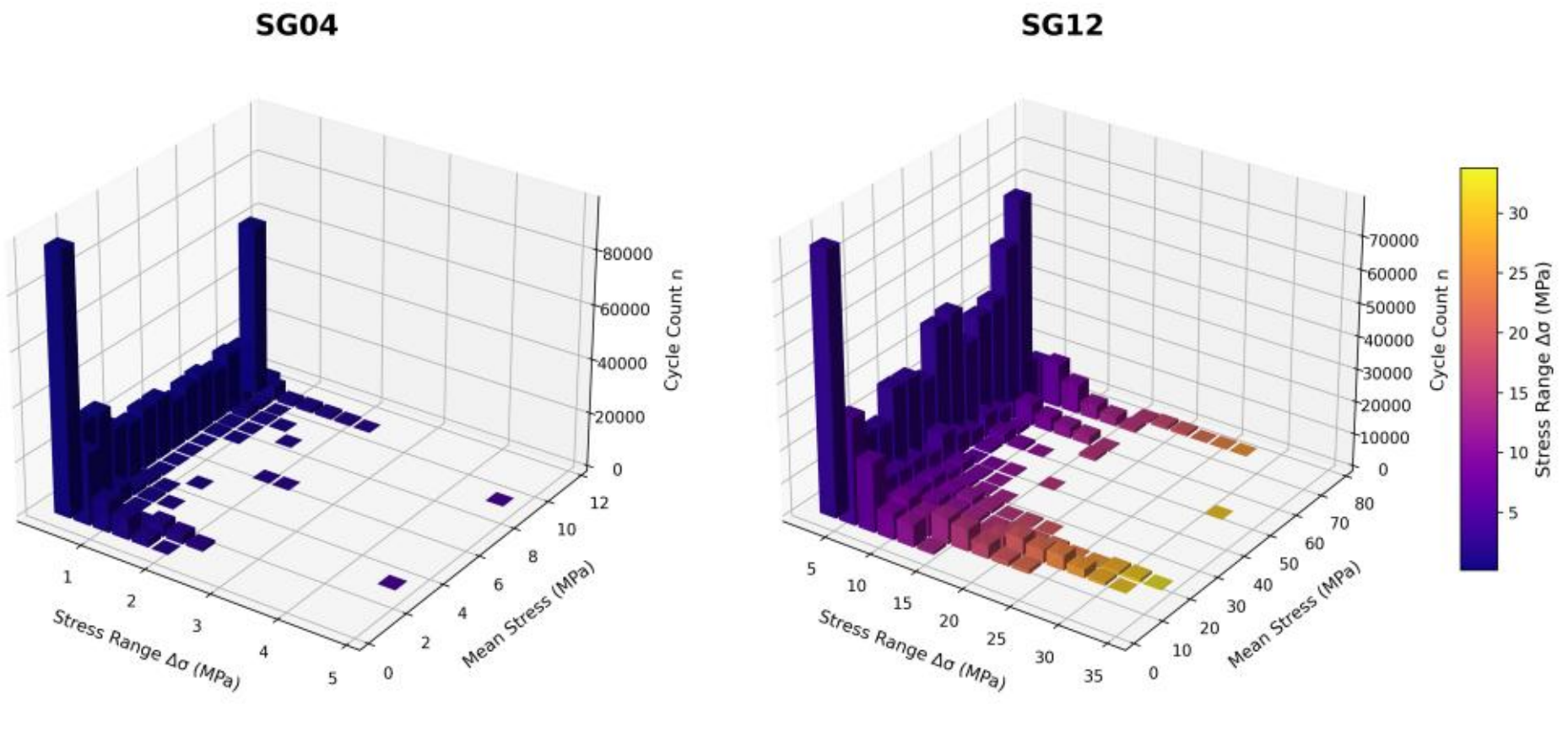

Figure 5. The rainflow count results for two representative sensors (SG04 and SG12).

Table 1. The probability models for the fatigue reliability analysis.

| Variable | Distribution | Mean value | Standard deviation |
|---|---|---|---|
| $\alpha_0(mm)$ | exponential | 1 | 1 |
| $\alpha_{th}(mm)$ | normal | 15 | 3 |
| $m$ | normal | 3 | 0.1 |
| $C_m$ | log-normal | $1.585\times10^{-13}$ | 0.4 |
| $v$ | constant | According to the cycle count | |

Table 2. Reference values of cost model parameters.

| Parameter | Bridge | | Reference |
|---|---|---|---|
| | The bridge in Sweden | The bridge in China | |
| $C_1$ | 5000€/m2 | 2000€/m2 | [55] |
| $C_2$ | 1,670,000€/person | 1,670,000€/person | [56] |
| $C_3$ | 10€/km | 0.50€/km | [57] |
| $C_4$ | 20€/km | 1.10€/km | [57] |
| $C_5$ | 15€/h | 5.7€/h | [58] |
| $C_6$ | 50€/h | 38€/h | [58] |
| $L_b$ | 42m | 1288m | - |
| $W_b$ | 5.35m | 37.2m | - |
| $X$ | 10 | 100 | - |
| $A$ | 10/day | 40000/day | - |
| $D_L$ | 50km | 30km | - |
| $T_a$ | 30% | 30% | - |
| $d$ | 365 | 365 | - |
| $O$ | 100 | 2 | - |
| $S_d$ | 100 km/h | 50 km/h | - |
| $\gamma$ | 0.9 | 0.9 | - |

### 3.1.2 Pre-posterior analysis and threshold optimization

This subsection determines the optimal warning threshold through pre-posterior analysis where the expected total cost associated with each candidate threshold is evaluated through the method in Section 2.3.3. Because the available data were collected under discrete experimental intervals (i.e., each bridge open event) rather than through continuous long-term monitoring, the pre-posterior analysis at decision epoch $\tau$ uses the five observation windows immediately following that epoch as the prospective observation inputs. It should be acknowledged that the five subsequent windows contain realized future observations that would not have been available at decision epoch $\tau$ . Therefore, this implementation serves as a retrospective demonstration of the sequential decision and threshold-updating procedure, rather than a strictly online prediction using only information available at each decision epoch.

Figure 6 shows the estimated expected total cost as a function of the candidate threshold for each sensor at the 58-th decision epoch, where the cost curve is evaluated exactly at the breakpoints generated by the shared scenario ensemble. The results indicate that the expected total cost and the optimal threshold vary among sensors. For sensors located far away from the damaged location (such as SG01-SG08), the results show that when the threshold is higher than $8.3 \times 10^{-4}$, the expected cost remains at a low level, which also means that the structural failure risk level remains at low and will not generate significant failure costs. In contrast, the optimal warning thresholds for sensors SG09 and SG10 near the damaged location are $6.201 \times 10^{-4}$ and $6.165 \times 10^{-4}$, respectively. When the threshold is higher, the cost induced by structural failure risk will be large, which means that the structure is subjected to a relatively high failure risk.

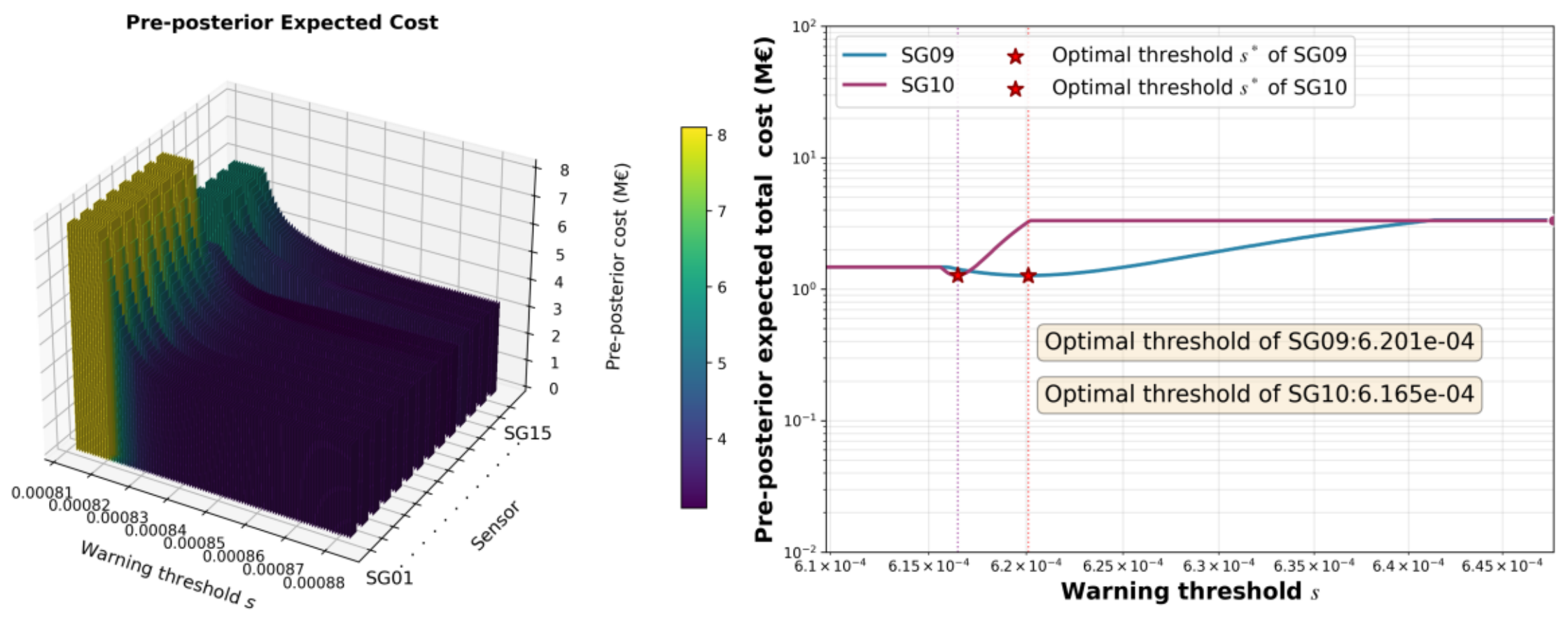

Figure 6. The expected cost as a function of the candidate threshold $s$ for each sensor. SG09 and SG10 are displayed separately.

Figure 7 presents the times at which the predicted failure risk exceeds the optimal threshold. It can be noted that, if the specified prediction period is relatively short, for instance, an interval spanning only two consecutive bridge opening events, or a horizon of a few days to one month, no warning would be issued at the current decision epoch under the optimized threshold. However, this does not imply that the proposed framework is unresponsive to damage, the optimal warning threshold is obviously changed due to the arrival of damage data. The detailed content is shown in the next section. Actually, the absence of an immediate warning is primarily attributed to the influence of the damage severity and the cost model on the warning decision. When the damage does not elevate the failure risk to a critical level, the warning is not triggered prematurely. From the perspective of the cost model, different proportions of direct and indirect costs produce different warning outcomes. To systematically investigate the effect of cost models, the cost ratio $r_c$ is defined:

$$r_c = \frac{C_{RB} + C_{LL}}{C_{RN} + C_{TL}} \tag{41}$$

Figure 8 illustrates how the optimal warning threshold and the corresponding threshold-violation instant of the failure risk vary with the ratio of direct to indirect costs. When direct and indirect costs are comparable, the optimal warning threshold remains relatively high and the threshold-violation instant is delayed; as the proportion of direct costs increases, the optimal threshold gradually decreases and the threshold-violation instant advances. This occurs because higher direct cost implies severe structural failure consequences, thereby lowering the tolerable risk level and prompting earlier intervention. Consequently, in the present case, if the direct cost proportion is substantially increased, i.e., the structural failure leads to catastrophic consequences, the warning decision would be triggered immediately.

Overall, the pre-posterior analysis yields a threshold that accounts for both current and future costs. Based on this threshold, proactive warning can achieve an optimal balance between early warning response costs and structural failure risk. A low warning threshold causes continuous cost from frequent warnings; consequently, raising the threshold is necessary to reduce the expected cost. However, when the threshold reaches a certain range, the failure cost becomes large, and a warning decision is needed. Conversely, an extremely high threshold exposes the structure to enormous failure risk costs. Therefore, an optimal balance exists that reduces warning response costs and maintains structural failure risk at a low level. This study is precisely aimed at finding this balance in a sequential manner.

In terms of computational efficiency, the pre-posterior analysis is accelerated by the shared-scenario threshold optimization algorithm in Section 2.3.3. In a conventional grid search implementation, the closed-loop scenario simulation is repeated for every candidate threshold, and the computation time therefore grows linearly with the number of candidate thresholds. In the proposed implementation, one ensemble of scenarios is generated once and shared by all candidate thresholds, and the threshold optimization is completed by a sorting operation followed by a single scan over the resulting breakpoints. The computation time is therefore governed mainly by the number of scenarios. Figure 9 compares the computation time of the two implementations. The computation time of the grid search implementation increases with both the number of scenarios and the number of candidate thresholds, whereas that of the shared-scenario implementation increases only with the number of scenarios and remains nearly constant as the number of candidate thresholds increases. Moreover, the shared-scenario implementation returns the exact minimizer over the admissible range rather than the best grid point, so the discretization error introduced by the grid search is also removed. These results also indicate that near real-time updating of warning thresholds is feasible.

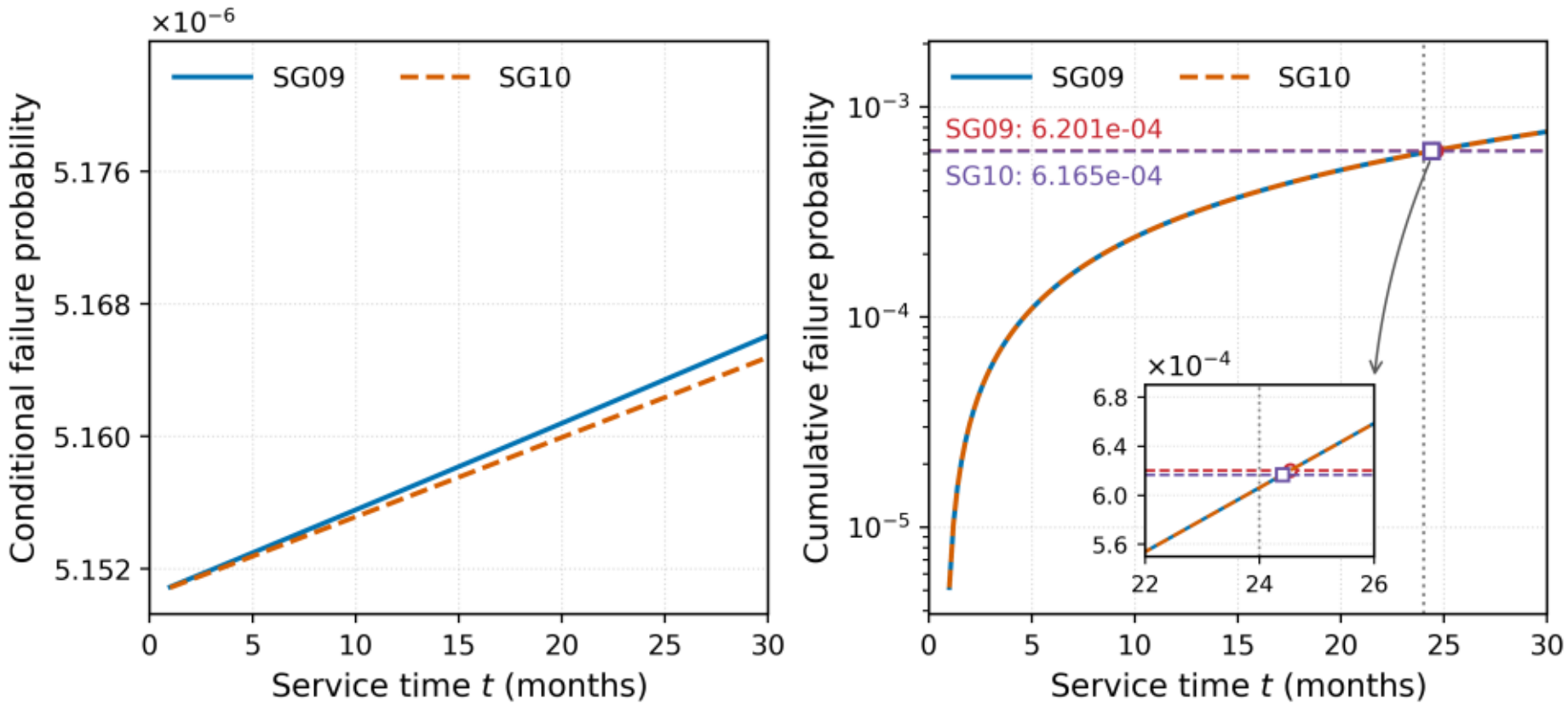

Figure 7. The times at which the failure risk level exceeds the optimal thresholds for sensors SG09 and SG10.

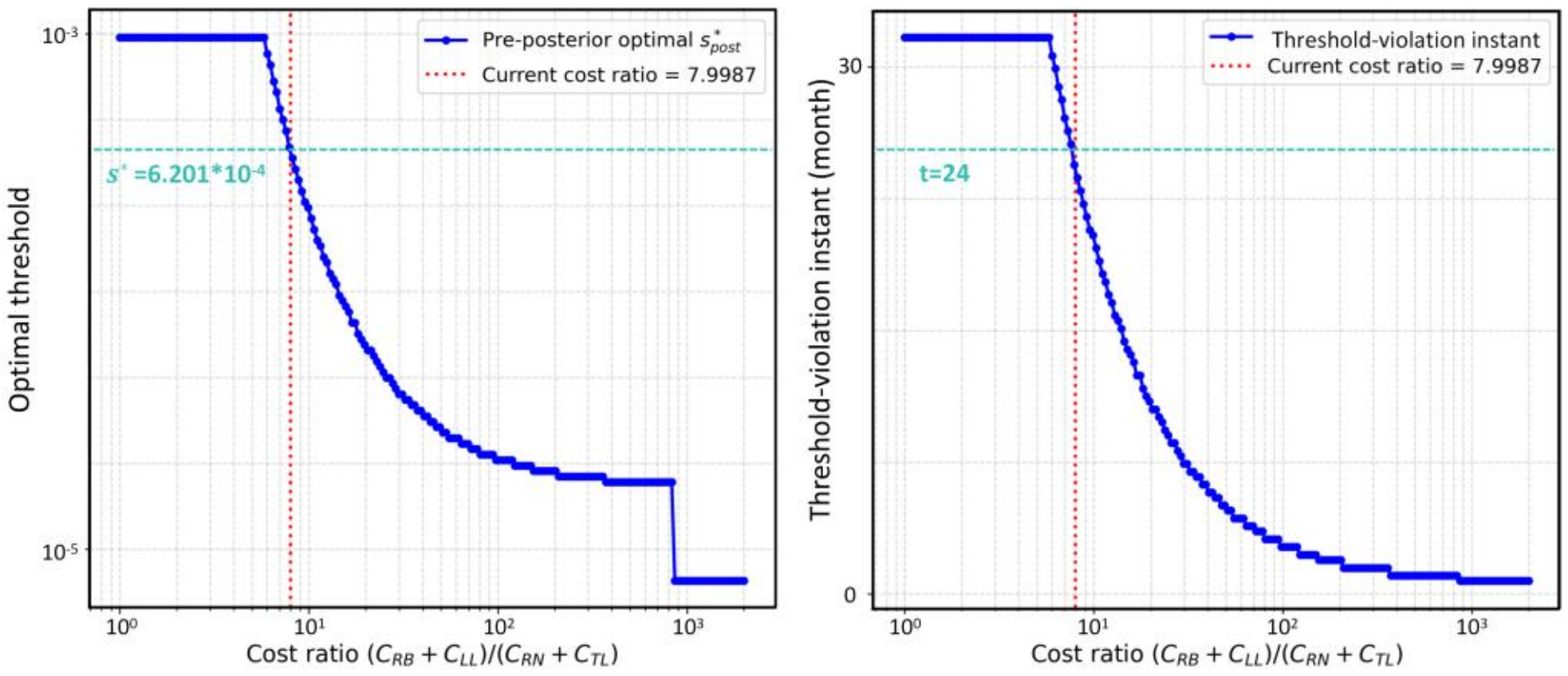

(a) (b)

Figure 8. The trend of optimal warning threshold of SG09 and threshold-violation instant when the ratio of direct and indirect costs changes. (a) Optimal warning threshold from pre-posterior analysis as a function of the cost ratio; (b) threshold-violation instant as a function of the cost ratio.

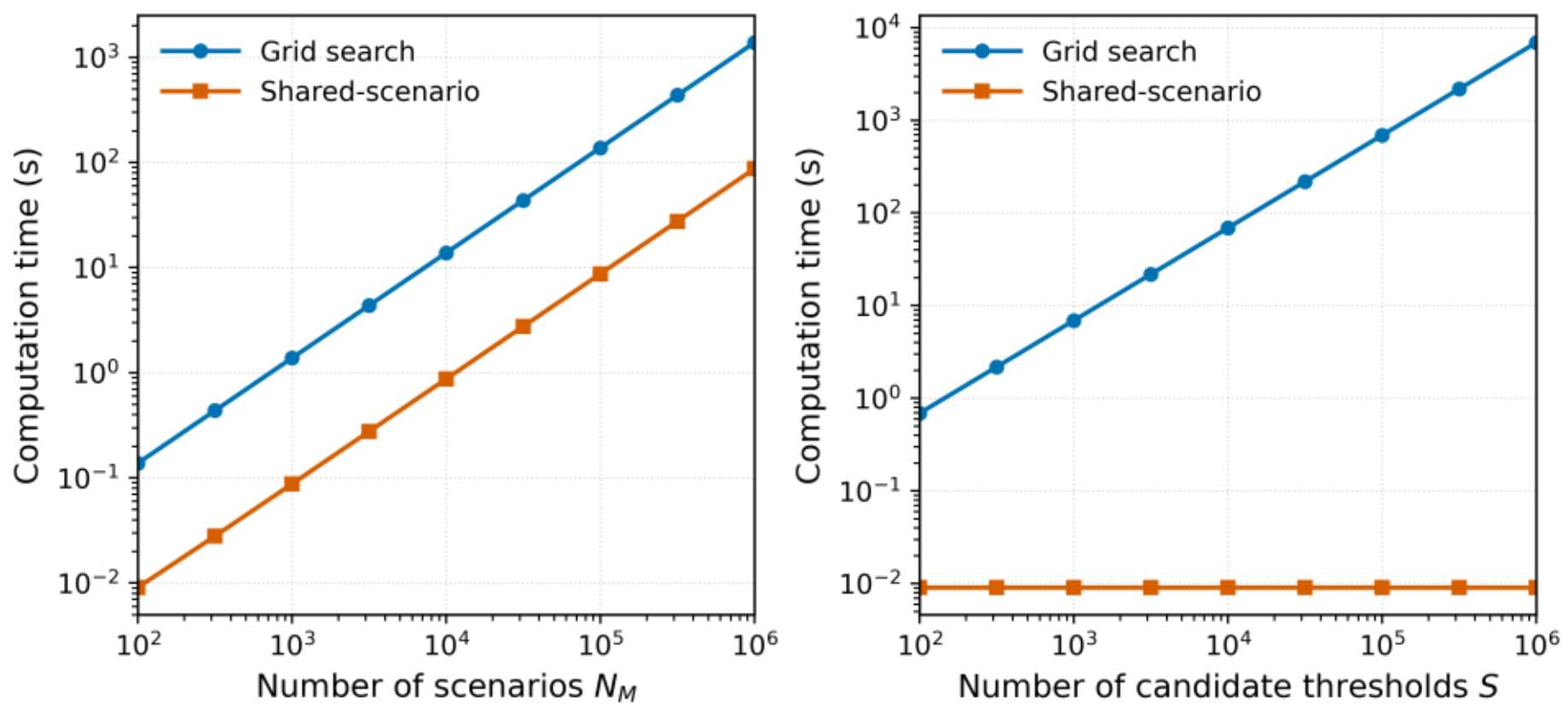

Figure 9. Comparison of computational efficiency between the grid search implementation and the shared-scenario implementation of the threshold optimization. (a) Computation time as a function of the number of scenarios with a fixed number of candidate thresholds. (b) Computation time as a function of the number of candidate thresholds with a fixed number of scenarios.

### 3.1.3 Adaptive updating of warning threshold

The adaptive updating capability of the warning threshold in the proposed framework is demonstrated through continuously acquired monitoring data windows, as shown in Figure 10. Specifically, following the order of bridge opening events, each monitoring data window is treated as new observations to update the belief state. At each decision epoch, the optimized threshold $s_\tau^*$ depends on both the updated belief state $b_\tau$ and the five prospective observation windows, and the warning threshold is re-optimized through the pre-posterior analysis. As the decision epoch advances, the belief state is also updated and the five window set shifts forward. Consequently, the threshold variations in Figure 10 reflect the combined effects of these two updating mechanisms and should not be attributed solely to the arrival of current observations.

Figure 10 illustrates the sequence of the optimal warning thresholds at each sensor. Sensors with similar results are plotted using the same line style and similar colors for visual grouping. The results demonstrate that the framework effectively responds to the damage information. The warning thresholds are automatically tightened as the structure deteriorates. If the warning threshold is not tightened and remains at the prior optimal value, the warning decision will be compromised when structural damage occurs, incurring unnecessary costs.

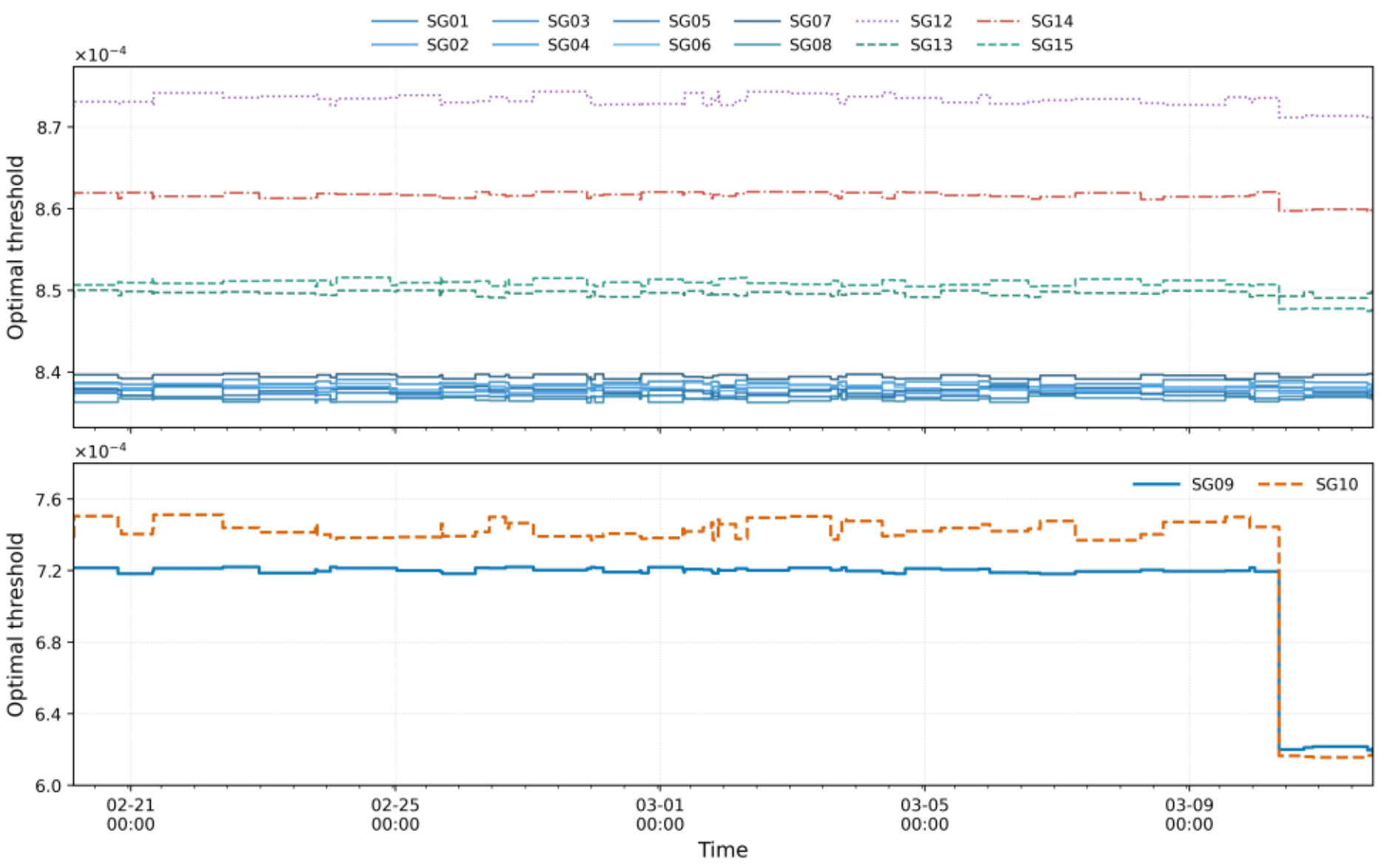

Figure 10. The sequence of the optimal warning thresholds for each sensor.

## 3.2 Early warning of stay-cable failure risk for a cable-stayed bridge in China

In this section, we will validate the proposed method using the dataset from an in-service cable-stayed bridge in China reported in [59]. The bridge is a six-lane cable-stayed bridge with double towers and double cable planes, as shown in Figure 11. The bridge has a span of length 1,288 m, featuring a steel box girder with a width of 37.2 m and a height of 3.2 m. "S/N" represents the cables on the south tower side and north tower side, respectively, and "A/J" denotes the cables on the bank side and riverside, respectively. The bridge was equipped with a comprehensive SHM system, including strain gauges, accelerometers and temperature sensors, which were deployed at critical structural components such as the main girder and cable anchors. The monitoring system continuously records structural responses under various operational and environmental conditions and has accumulated massive historical datasets that enable the probabilistic characterization of structural states.

Stay cables are critical components of cable-stayed bridges, and their fatigue failure or rupture may lead to catastrophic structural collapse. In the bridge considered in this study, the cable designated as SJS11 was found to be damaged during the monitoring period, the damage is rupture of wires (as shown in Figure 12). Starting from the year with available monitoring data, the damage was found by manual inspection in the sixth year. The monitoring data from two cable force sensors (SJS11 and SJS12) are compared to verify whether the proposed framework can make correct early warning decisions.

In this case study, the decision interval is set to one month. At each decision epoch, the failure risk over the upcoming year is evaluated (i.e., the risk horizon is twelve months), and a warning decision is made according to the warning threshold optimized through pre-posterior analysis. The monitoring data of the first available month are used to establish the initial belief state, which is continuously updated with the subsequently arriving data. Considering the cross-sectional area and elastic modulus of the cable, the raw cable force time histories are converted into stress ranges, which are then processed by rainflow counting to establish the distribution of the equivalent stress range. The stress time histories of cables SJS11 and SJS12 and their rainflow counting results on selected dates of the first monitoring period are shown in Figure 13. Fatigue reliability analysis is performed using the probability models listed in Table 3.

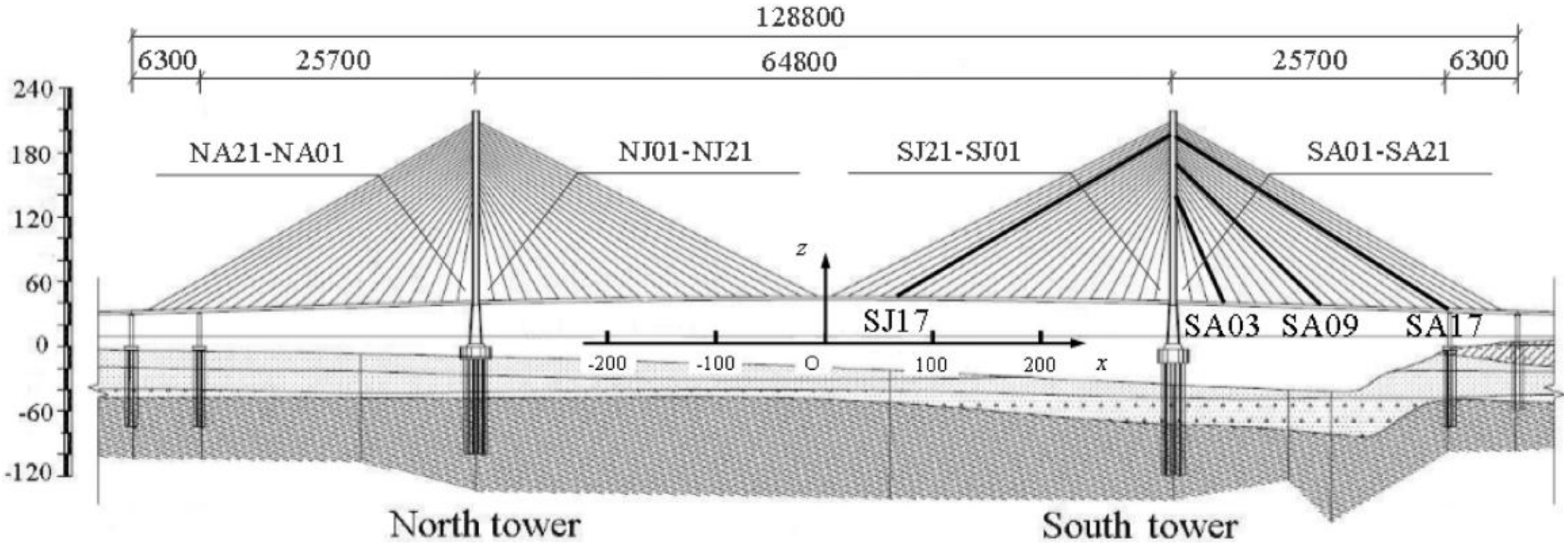

Figure 11 An in-service cable-stayed bridge in China reported in [59].

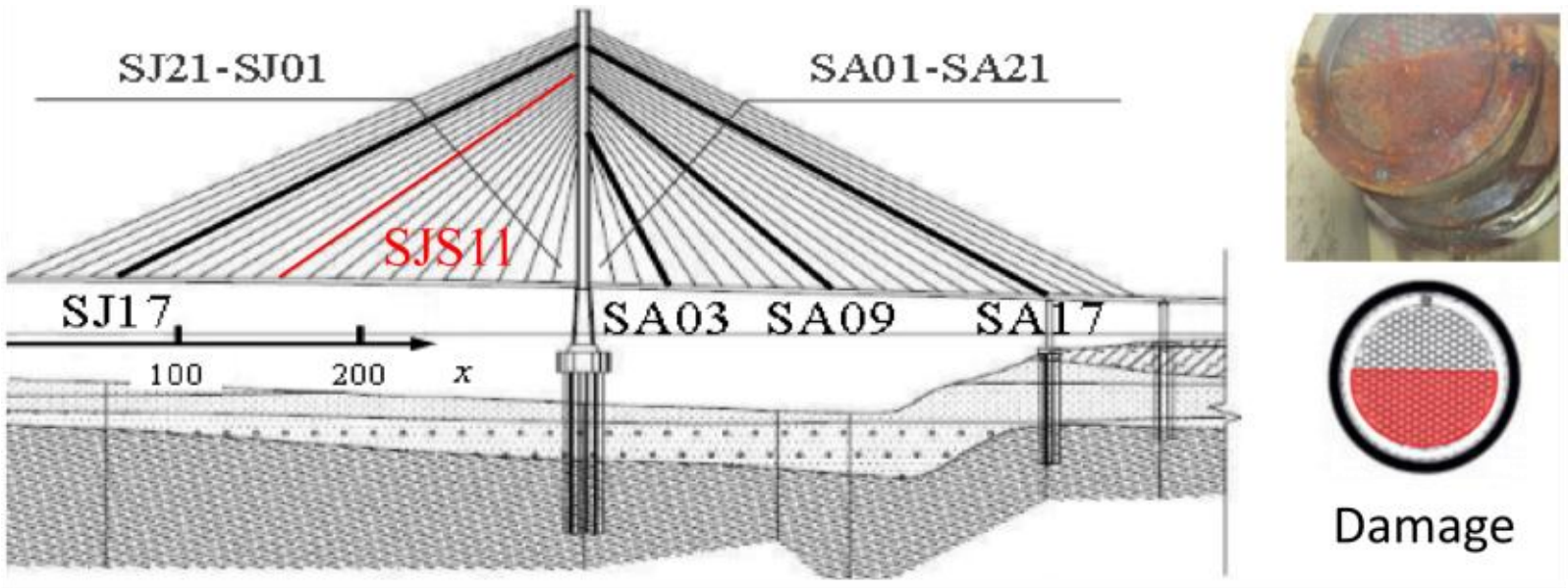

Figure 12. The damage occurred at SJS11.

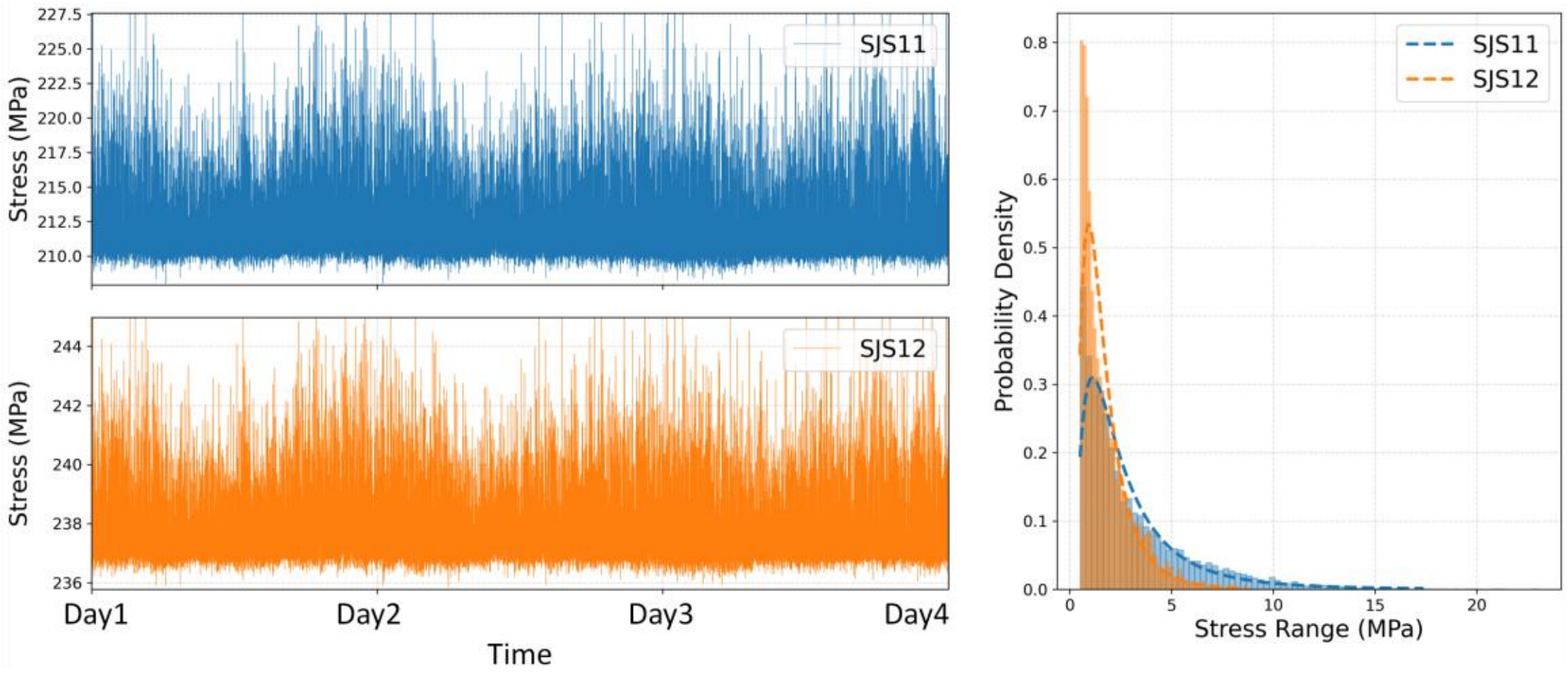

Figure 13. The raw data of cable stress time histories recorded on several representative days of the first monitoring period. SJS11 and SJS12 are selected as representative examples. The rainflow counting results of them are showed on the right side.

To demonstrate that the proposed framework can make correct early warning decisions, the analysis results at the 53$^{rd}$ decision epoch are presented. At all preceding decision epochs, the optimal decision is not to trigger the warning, as the predicted failure risk within one year would not exceed the threshold; at the 53$^{rd}$ decision epoch, however, the framework anticipates the failure risk of the cable in the sixth year and an early warning decision is made accordingly.

Specifically, at the 53$^{rd}$ decision epoch, the warning threshold is optimized by the pre-posterior analysis, which uses the scenario-based method to simulate all possible early warning decision costs. The Bayesian DLM is employed to forecast the prospective monitoring data over the upcoming year, i.e., $n = 12$. Predictions beyond the one-year horizon are not used, since the predictive uncertainty grows with the lead time and the discount factor down-weights the distant costs, making their contribution to the expected total cost negligible. Figure 14 presents the optimal warning threshold obtained through pre-posterior analysis.

Based on the optimal threshold, the warning decision can be made for SJS11, as shown in Figure 15. The predicted failure risk over the upcoming year exceeds the optimized threshold at the 53rd decision epoch, so that an early warning is issued. This indicates that, by forecasting the prospective monitoring data with the Bayesian DLM and evaluating the decision consequences through the scenario-based analysis, the framework anticipates the risk elevation and triggers the warning before the damage occurs, which is found to be in the upcoming year. In contrast, cable SJS12 maintains a consistently high safety level, and no high-risk condition arises within the following 60 months. Overall, the proposed method successfully triggers an early warning of the bridge failure risk. By jointly considering structural deterioration and warning response costs in the pre-posterior analysis, a cost-effective early warning decision is achieved.

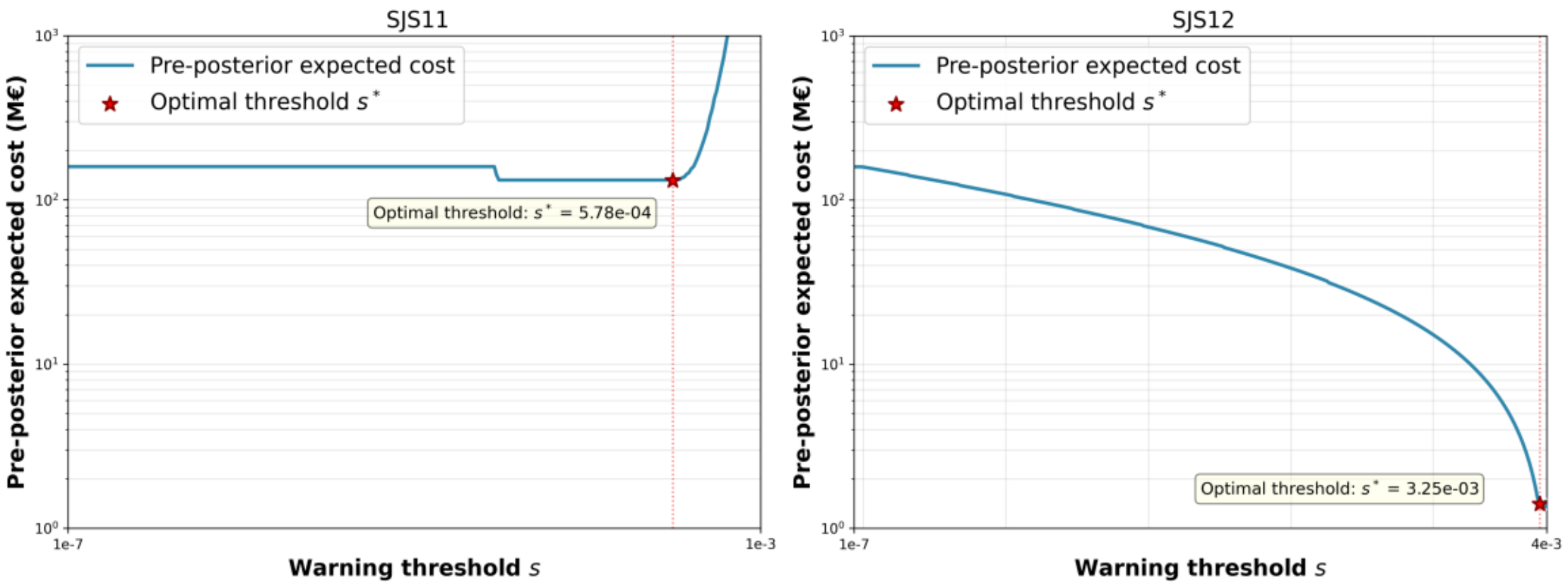

Figure 14. The optimal warning threshold obtained through pre-posterior analysis.

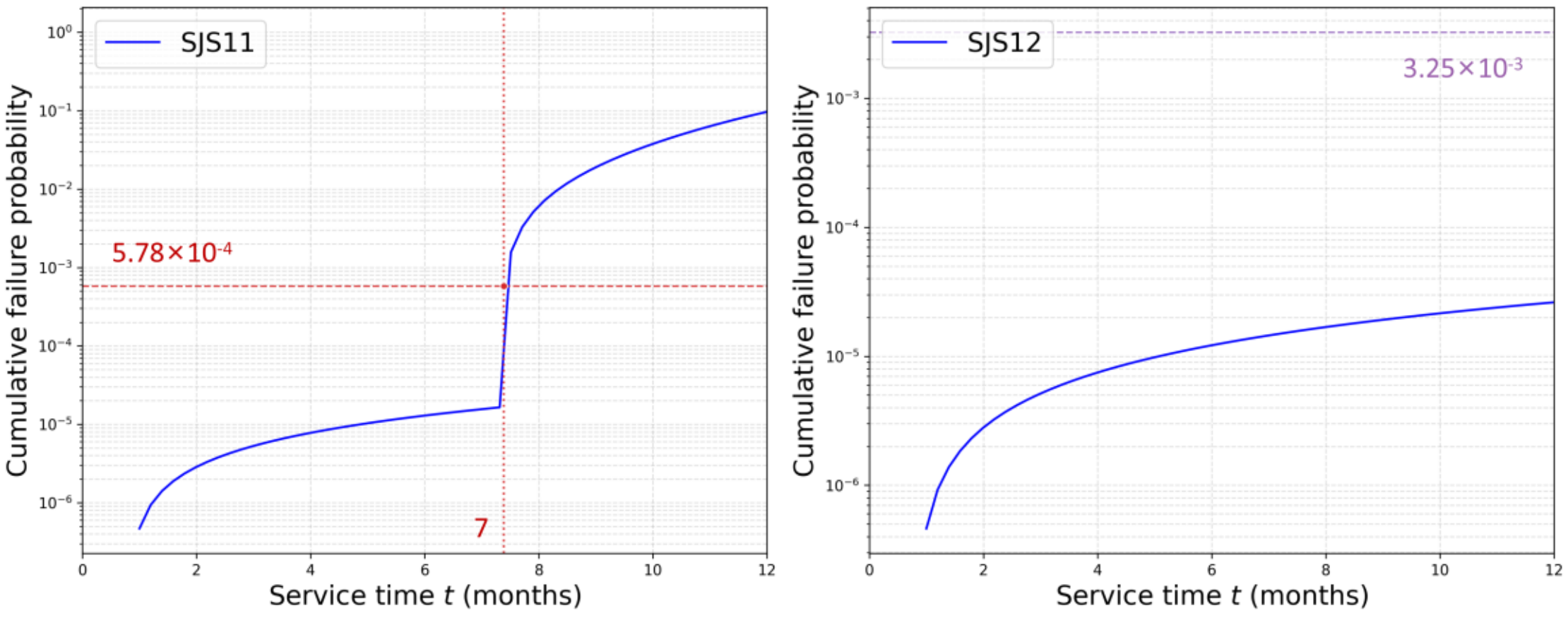

Figure 15. The threshold-violation instant decided by the predicted risk and warning threshold. The warning action of SJS11 should be triggered at the current decision epoch where the predicted failure risk exceeds the optimal threshold, and SJS12 will not trigger any warning action.

Table 3. The probability models for the fatigue reliability analysis of cables.

| Variable | Distribution | Mean value | Standard deviation |
|---|---|---|---|
| $\alpha_0(mm)$ | exponential | 0.05 | 0.0025 |
| $\alpha_{th}(mm)$ | normal | 2.5 | 1 |
| $m$ | normal | 3 | 0.1 |
| $C_m$ | log-normal | $5\times10^{-13}$ | 0.5 |
| $v$ | constant | According to the cycle count | |

### 3.3 Comparison with a fixed-threshold policy

The proposed framework is compared with the most relevant alternative, i.e., a warning scheme in which the risk threshold is optimized once and then held fixed throughout the service period. This alternative shares exactly the same information, deterioration model, observation model and cost structure as the proposed framework, and differs only in whether the threshold is re-optimized at each decision epoch. The quantitative comparison is enabled by the scenario-based evaluation, which provides the expected total cost of every candidate fixed threshold. According to Figure 6 and Figure 10, the threshold fixed at the first decision epoch yields an expected total cost of 3.31 million euros, whereas the adaptively re-optimized policy achieves 1.287 million euros. Moreover, with the fixed threshold, the warnings would be issued at least 50 months later, which is much later than the 24th month of the proposed framework (Figure 7). At that time, the predicted failure risk had already reached to a high level. These results indicate that the advantage of the adaptive threshold is not merely conceptual but translates into measurable cost savings and earlier warnings. This observation is also consistent with the theoretical guarantee stated in **Appendix C**. .

A fixed risk threshold would be cost-effective only under stationary conditions. In practice, the uncertainty in the structural state estimate evolves as monitoring data accumulate, and the cost of warning responses varies over time. The warning threshold at which issuing a warning minimizes the expected total cost therefore shifts continuously. A fixed threshold cannot track this shift: if set too low, it triggers premature warnings and unnecessary closure costs in the early service period when the failure risk is still low; if set too high, it delays the warning when deterioration accelerates and severe consequences become imminent. The proposed framework re-optimizes the threshold at every decision epoch using the latest information, thereby tracking the evolving cost–risk balance.

### 3.4 Discussions

According to the methodologies and the case studies, the advantages of the proposed framework include: (i) it achieves the trade-off between structural failure risk and warning response costs by minimizing the expected total cost, whereas traditional methods ignore the economic consequences of decisions; (ii) it is proactive, since future monitoring data are predicted and the decision is optimized over these predictions, whereas traditional methods are reactive and issue delayed alarms; and (iii) it is adaptive, the warning threshold are updated at every decision epoch using the latest information, so that the warning policy tracks the evolving uncertainties and time-varying costs, whereas traditional methods rely on a fixed threshold. However, in practical implementations, some limitations should also be discussed.

### 3.4.1 Model assumptions and limitations

The assumptions underlying equations (15) and (16) include the stationary $\mu$ and independent stress cycles. Under nonstationary or strongly correlated conditions, these assumptions may influence the updating behavior. For example, accumulating all historical observations may slow adaptation to abrupt damage-induced changes in $\mu$. In addition, the posterior variance in Eq. (16) decreases whenever new data are added, even if those data are anomalous, because it quantifies only the uncertainty in estimating the stationary parameter $\mu$ and does not account for additional uncertainty arising from unexpected structural changes. Moreover, positive correlation among rainflow cycles may also lead to underestimated sampling variability and an overconfident posterior belief.

However, for problems involving gradually evolving damage and monitoring windows containing sufficient cycles (such as the fatigue deterioration in the case studies), these assumptions provide a reasonable approximation and can support closed-form and computationally efficient Bayesian updates. For more complex scenarios involving abrupt damage or strongly correlated cycles, several extensions can be considered. A time-varying state model or a forgetting factor could assign greater influence to recent observations and improve the response of the belief update to changes in $\mu$. Adaptive model discrepancy terms could allow anomalous observations to introduce additional uncertainty rather than being treated as ordinary measurements. Furthermore, effective sample sizes estimated from the autocorrelation of the monitoring series, or covariance-aware likelihoods, could be used to represent cycle dependence and avoid overconfident posterior estimates. These extensions provide useful directions for future research.

### 3.4.2 Influence of the cost model

According to the computation logic of the presented framework, the early warning results depend heavily on the cost model. The optimal threshold is essentially the point where the costs of an unnecessary response equals the failure risk costs. Therefore, the weighting of reconstruction costs, life-loss costs, and traffic disruption costs directly determines where the threshold falls. If direct costs are assigned a high value, the framework will favor a stricter threshold to avoid catastrophic failure even if this produces more false alarms. If traffic disruption costs dominate, the framework will prefer a more lenient threshold to reduce unnecessary closures. Because these costs vary by region, the threshold obtained in this study should be viewed as conditional on the assumed cost structure rather than a universal recommendation. It is worth noting that structural failure costs typically far exceed the indirect costs associated with traffic disruption and time delays. Consequently, the proposed framework triggers warning actions while the structural failure risk remains relatively low, thereby preventing the risk from escalating to a high level.

## 4. Conclusion

This study develops a Bayesian sequential decision-making framework for early warning of bridge failure risk in time-varying systems. Unlike traditional methods that issue alarms reactively once a monitoring indicator exceeds an empirically predefined static threshold, the proposed framework treats early warning as a sequence of decision-making problems under uncertainty throughout the service life. The framework integrates a probabilistic state transition model, an observation prediction model, cost models of early warning decisions, and a threshold-based decision strategy into a unified closed loop. Within this loop, structural state updating, prospective observation forecasting, warning threshold optimization, and warning decision-making are sequentially executed at each decision epoch. The prospective observations are forecast by a Bayesian dynamic linear model and on the basis of these forecasts, pre-posterior analysis evaluates the expected total cost of all candidate thresholds from one shared scenario ensemble. Then the optimal threshold is determined by minimizing the expected total cost, which achieves an explicit balance between failure risk and warning response costs. The threshold is optimized in a rolling manner as new monitoring data arrive, so that the warning strategy adapts to the evolving uncertainties.

The proposed methodology is validated using both a dataset from a steel bridge in Sweden and a long-term monitoring dataset from an in-service cable-stayed bridge in China. These two cases cover typical scenarios that lead to increased structural failure risk, including steel truss member fracture and stay cable rupture. Results from the first case demonstrate that the proposed framework strikes a balance between the structural failure risk and the warning response costs. Although structural damage elevates the failure risk level and advances the time at which it exceeds the warning threshold, triggering a warning without delay may not be the economically optimal decision when the damage severity or the proportion of direct costs is relatively low. Conversely, if the damage is severe enough to substantially elevate the failure risk level, or if direct costs dominate, a warning decision is made immediately. The sensitivity analysis on the cost model shows that the warning timing advances adaptively as failure consequences outweigh response costs. For the second case, the study results indicate that under the current parameter settings, the healthy cable maintains stable performance across the analyzed time intervals, with the optimal policy being continuous normal operation. The damaged cable triggers an early warning decision before the damage elevates the structural failure risk to a high level. Although the optimal warning thresholds are sensitive to the cost model, the proposed method generally triggers warning actions before the structural failure risk escalates to high levels under typical configurations where failure costs substantially outweigh indirect costs, thereby preventing catastrophic consequences. In addition, the adaptive updating mechanism of this framework enables timely refinement of warning thresholds as new observations are acquired, ensuring that early warning decisions remain aligned with the evolving structural state.

Finally, future research will leverage large language models and intelligent agents to integrate heterogeneous data sources, such as video streams, textual records, and multi-type sensor measurements, enabling more comprehensive structural state characterization and autonomous adjustment of warning strategies without constant human intervention.

## 5. Acknowledgement

This study was supported by National Key Research and Development Program of China [Grant No. 2021YFF0501003], Jinan City Science and Technology Program Project [Grant No.

202428059], Guizhou Provincial Major Science and Technology Special Project [Project No. Guizhou Major Science and Technology Contract (2025) 033] and Guizhou Provincial Major Science and Technology Special Project [Project No. Guizhou Scientific Research Platform Contract (2025) 013]

## Appendix A. An example of cost models for bridge early warning

In reference [60], the costs are classified into four categories: bridge reconstruction costs $C_{RB}$, life loss costs $C_{LL}$, bridge running costs $C_{RN}$ and time loss costs $C_{TL}$. Among these, reconstruction expenses and the costs related to life loss are direct costs produced by immediate consequences of bridge failure. Conversely, bridge running expenses and time-loss costs are categorized as indirect costs, arising from either the management interventions after failure event or traffic control measures. According to reference [61], these costs can be computed by the following equations:

$$C_{RB} = C_1 W_b L_b \tag{A-1}$$

$$C_{LL} = C_2 N_{LL} \tag{A-2}$$

$$C_{RN} = [C_3(1 - T_a) + C_4 T_a] D_L A d \tag{A-3}$$

$$C_{TL} = [C_5 O(1 - T_a) + C_6 T_a] \frac{D_L A d}{S_d} \tag{A-4}$$

where $C_1$ represents the unit rebuilding cost; $W_b$ and $L_b$ are the width and the length of the bridge, respectively; $C_2$ is the cost for each life lost; $N_{LL}$ is the life loss number; $C_3$ and $C_4$ represent the cost of cars and trucks, respectively; $T_a$ is the average percentage of trucks in a whole day; $D_L$ is the detour length; $A$ is the number of daily vehicles; $d$ denotes duration of the detour; $C_5$ represent the value of time for each car passenger; $O$ is the average car occupancy rate; $C_6$ is the value of time for trucks; $S_d$ is the average speed during the detour.

## Appendix B. An example of structural deterioration model

This study adopts a fatigue reliability analysis framework to probabilistically characterize the time-varying evolution of structural states based on fracture mechanics theory. Bridge structures often experience cyclic stress variations induced by repeatedly applied forces, which lead to progressive accumulation of fatigue damage and the initiation and propagation of fatigue cracks. When these cracks extend to critical lengths, the service performance of the structure becomes compromised. For instance, the flexural capacity of girders may degrade substantially, which makes the bridge vulnerable to damage caused by fracture during operation and leads to serious economic losses. The core assumption of the theory is that fatigue cracks spread from initial defects and propagate under cyclic loads until a critical crack width or depth is reached. According to linear elastic fracture mechanics (LEFM) [62], the range of the stress intensity factor governs the rate of crack propagation. The Paris-Erdogan law describes the one-dimensional fatigue crack extension as follows:

$$\frac{d\alpha}{dN} = C_m(\Delta K)^m \tag{B-1}$$

$$\Delta K = G\Delta\sigma\sqrt{\pi\alpha} \tag{B-2}$$

where $\frac{d\alpha}{dN}$ is the crack growth rate; $C_m$ and $m$ are material constants determined by fatigue tests; $\Delta K$ is the stress intensity factor range; $G$ is the crack geometric correction factor, which is specified to be 1 in this study for simplification. According to reference [63], the following equation can be derived:

$$\alpha(t) = \left(\alpha_0^{\frac{2-m}{2}} + \frac{2-m}{2}C_m\left(\sqrt{\pi}\Delta\sigma(t)\right)^m vt\right)^{\frac{2}{2-m}} \tag{B-3}$$

where $\alpha_0$ is the initial fatigue crack depth; $v$ is the average number of load cycles, which can be determined according to the rainflow count analysis; $\alpha(t)$ represents the fatigue crack depth at time $t$. The fatigue limit state function can be defined as:

$$g(t) = \alpha_{th} - \alpha(t) \tag{B-4}$$

where $\alpha_{th}$ is the limit threshold of fatigue crack. The structural failure probability and reliability are calculated by the HL-RF algorithm [64]:

$$P_f(t) = P(g(t) = \alpha_{th} - \alpha(t) < 0) = \Phi(-\beta) \tag{B-5}$$

While the above equation gives the failure probability at a specific time, the assessment of structural performance over a design lifetime needs to compute the cumulative failure probability within a discrete-time framework. If the time horizon $[0, t_n]$ is partitioned into $n$ discrete intervals $\Delta t_i (i = 1,2,\ldots,n)$, and the conditional probability of failure within each interval is presented by $P_f(t_i)$, then the cumulative failure probability up to time $t_n$ can be expressed as:

$$P_c(t_i) = 1 - \prod_{j=1}^{i}[1 - P_f\left(t_j\right)] \tag{B-6}$$

The failure probability and cumulative failure probability provide the assessment of structural safety performance over time. Based on them, the hazard function $h(t_i)$ can be define:

$$h(t_i) = P_c[T \in [t_{i-1}, t_i]|T > t_{i-1}] = \frac{P_c(t_i) - P_c(t_{i-1})}{1 - P_c(t_{i-1})} \tag{B-7}$$

This function represents the conditional probability that a structure fails during the $i$-th interval, given that it has successfully survived all previous intervals up to time $t_{i-1}$.

## Appendix C. Policy-improvement property of the adaptive threshold policy

We first present the assumptions for this part. A proposition, the proof of this proposition and some remarks on its interpretation are then illustrated. The proposition presented in this appendix adapts the standard rollout/policy-improvement property of dynamic programming to the threshold-based warning policies considered in this study. It is not intended as a new general result in dynamic programming. Rather, it clarifies the assumptions under which adaptive threshold re-optimization improves upon a fixed base threshold within the proposed early-warning framework. Specifically, the guarantee is established under the following assumptions:

(A1) The service horizon is finite and the one-step costs are bounded;

(A2) The expected total cost of any candidate policy is evaluated without Monte Carlo estimation error. The influence of the estimation error is examined in the remarks;

(A3) The state transition model, the observation prediction model, the cost models, and the risk prediction model, are identical at all decision epochs, and the belief is updated by Bayesian filtering with fixed model parameters;

(A4) The admissible threshold range is identical at all decision epochs, so that any fixed threshold remains feasible at every optimization process.

(A5) Upon the arrival of a new observation, the belief state is updated to the exact Bayesian posterior, so that the future cost averaged over the predicted observations equals the true expected future cost.

**Proposition 1**. *Under assumptions (A1) to (A5), for any decision epoch* $\tau$ *and any belief state* $b_\tau$*, the expected total cost of the adaptive threshold policy* $\pi^{\mathrm{ad}}$ *is no greater than that of any fixed threshold policy, i.e.,*

$$V_\tau^{\mathrm{ad}}(b_\tau) \le J_\tau(s \mid b_\tau), \forall s \in \mathcal{S} \tag{C-1}$$

*Proof.*

For each epoch $k$ and each belief state $b_k$, the expected total cost of a fixed threshold $s$ over the remaining horizon satisfies:

$$J_k(s \mid b_k) = \begin{cases} \bar{C}_{w,k}(b_k), R_k(b_k) \ge s \\ \bar{q}_k(b_k)C_F + \gamma\mathbb{E}[J_{k+1}(s \mid b_{k+1})], R_k(b_k) < s \end{cases} \tag{C-2}$$

The terminal condition is $J_{T+1} = 0$, where $T$ is the terminal decision epoch defined in section 2.1. The optimized cost value can be obtained by:

$$V_k(b_k) = \min_{s \in \mathcal{S}} J_k(s \mid b_k) \tag{C-3}$$

Since the minimization is taken over a set that contains $s$, it holds that $V_k(b_k) \le J_\tau(s \mid b_k)$ for every $s \in \mathcal{S}$. The proof need to show that the adaptive policy attains a cost no greater than $V_k(b_k)$, which is carried out by backward induction over $k = T, T-1, \dots, \tau$. At the terminal epoch $T$, only one decision remains, and the adaptive policy selects the threshold that minimizes the cost, i.e., $V_T^{\mathrm{ad}}(b_T) = V_T(b_T)$. Assume now that $V_{k+1}^{\mathrm{ad}}(b) \le V_{k+1}(b)$ holds for every belief $b$. At epoch $k$, let $s^*$ denote a minimizer of $J_k(\cdot \mid b_k)$, and note that the adaptive policy applies $s^*$ at epoch $k$. If $R_k(b_k) \ge s^*$, the adaptive policy issues a warning and terminates, so $V_k^{\mathrm{ad}}(b_k) = \bar{C}_{w,k}(b_k) = J_k(s^* \mid b_k) = V_k(b_k)$. If $R_k(b_k) < s^*$, both the adaptive policy and the fixed policy with threshold $s^*$ continue operation at epoch $k$, incur the same expected one-step cost, and face the same predictive distribution of the next observation, and therefore:

$$\begin{aligned} V_k^{\mathrm{ad}}(b_k) &= \bar{q}_k(b_k)C_F + \gamma\mathbb{E}\left[V_{k+1}^{\mathrm{ad}}(b_{k+1})\right] \\ &\le \bar{q}_k(b_k)C_F + \gamma\mathbb{E}[V_{k+1}(b_{k+1})] \\ &\le \bar{q}_k(b_k)C_F + \gamma\mathbb{E}[J_{k+1}(s^* \mid b_{k+1})] \\ &= J_k(s^* \mid b_k) = V_k(b_k) \end{aligned} \tag{C-4}$$

where the first inequality follows from the assumptions, and the second follows from $V_{k+1}(b) \le J_k(s^* \mid b)$. Combining the two cases yields $V_k^{\mathrm{ad}}(b_k) \le V_k(b_k)$, which completes the proof.

Some discussions of the proof should be presented. As stated in the assumptions,

the proposition is proved under exact evaluation. In the numerical implementation, the expected costs are estimated by the shared-scenario evaluation with $N_M$ scenarios. Within a single optimization at the current epoch, the empirical cost of the optimized threshold is no greater than that of any other candidate threshold evaluated on the same scenario ensemble, i.e. the inequality holds. However, different scenario ensembles may be generated when the threshold is re-optimized at subsequent decision epochs. Finite sample estimates are thus affected by sampling error and by the selection bias introduced when minimizing an estimated cost function, and the empirical costs may deviate from the ordering established under exact evaluation. Increasing $N_M$, repeating the simulations, or evaluating the resulting policies on independent validation ensembles can be used to examine the stability of the optimized thresholds and estimated policy costs. In addition, the proof relies on the fixed model premises, as stated in (A3). When the hyper-parameters of the related models are updated between epochs, the comparison at each epoch is conditioned on the latest premises, and the proposition still holds.